\documentstyle[prd,aps,floats,tabularx,epsfig,comment,subfigure]{revtex}

\tighten

\newcommand{\ETAL}{{\it et al.}}
\newcommand{\be}{\begin{equation}}
\newcommand{\ee}{\end{equation}}
\newcommand{\ba}{\begin{eqnarray}}
\newcommand{\ea}{\end{eqnarray}}
\newcommand{\hld}{\hspace{0.25cm}\cdots}
\newcommand{\fns}{\footnotesize}
\newcommand{\fnsc}{\scriptsize}
\newcommand{\mbs}[1]{\mbox{\small #1}}
\newcommand{\la}{\langle}
\newcommand{\ra}{\rangle}
\newcommand{\mdcs}[2]{\langle \mbox{\scriptsize#1,#2} \rangle}

\newcommand{\fiso}{f_{\mbox{\fnsc ISO}}}
\newcommand{\ziso}{z_{\mbox{\fnsc ISO}}}

\def\gtorder{\mathrel{\raise.3ex\hbox{$>$}\mkern-14mu
             \lower0.6ex\hbox{$\sim$}}}
\def\ltorder{\mathrel{\raise.3ex\hbox{$<$}\mkern-14mu
             \lower0.6ex\hbox{$\sim$}}}

\begin{document}

\draft
\preprint{\
\begin{tabular}{rr}
&
\end{tabular}
}
\twocolumn[\hsize\textwidth\columnwidth\hsize\csname@twocolumnfalse\endcsname
\title{Constraints on isocurvature models from the WMAP first-year data}
\author{K.~Moodley$^{1,2}$, M.~Bucher$^3$, J.~Dunkley$^2$, P.~G.~Ferreira$^2$,
 C.~Skordis$^2$}
\address{
$^1$School of Mathematical Sciences,
University of KwaZulu-Natal, Durban, 4041, South Africa\\
$^2$Astrophysics, University of Oxford,
Denys Wilkinson Building, Keble Road, \\ Oxford OX1 3RH, United Kingdom \\
$^3$DAMTP, Centre for Mathematical Sciences, University of Cambridge,
Wilberforce Road, \\ Cambridge CB3 0WA, United Kingdom }
\maketitle

\begin{abstract} 
We investigate the constraints imposed by the first-year WMAP CMB data
extended to higher multipole by data from ACBAR, BOOMERANG, CBI and the VSA
and by the LSS data from the 2dF galaxy redshift survey on the 
possible amplitude 
of primordial isocurvature modes. A flat universe with CDM and $\Lambda $ 
is assumed, and the baryon, CDM (CI), and neutrino density (NID) 
and velocity (NIV) isocurvature modes are considered. Constraints 
on the allowed 
isocurvature contributions are established from the data for
various combinations of the adiabatic mode and one, two, and three 
isocurvature 
modes, with intermode cross-correlations allowed. Since baryon and CDM 
isocurvature are observationally 
virtually indistinguishable, these modes are not considered 
separately. We find that when just a single isocurvature mode is
added, the present data allows an isocurvature
fraction as large as $13\pm6$, $7\pm4$, and $13\pm7$ percent for
adiabatic plus the CI, NID, and NIV modes, respectively. When
two isocurvature modes plus the adiabatic mode and cross-correlations 
are allowed, these percentages rise to $47\pm16$, $34\pm12$, and 
$44\pm12$ for the 
combinations CI+NID, CI+NIV, and NID+NIV, respectively. Finally,
when all three isocurvature modes and cross-correlations are allowed,
the admissible isocurvature fraction rises to $57\pm9$ per cent. 
The sensitivity 
of the results to the choice
of prior probability distribution is examined.
\end{abstract}

\date{\today}
\pacs{PACS Numbers : 98.80.-k}]
\renewcommand{\thefootnote}{\arabic{footnote}} \setcounter{footnote}{0}
\noindent


\section{Introduction}

Although the current observational data on the cosmic 
microwave background (CMB) anisotropy and large-scale
structure (LSS) seem consistent with a spatially 
flat universe with a cosmological constant ($\Lambda $)
filled with ordinary matter (baryons, leptons, photons,
and neutrinos) and a cold dark matter (CDM) component
with a primordial spectrum of Gaussian adiabatic density 
perturbations described by a monomial power law 
[i.e., $P(k)\propto k^{n_s}$], it is of interest to consider whether 
models having an appreciable isocurvature component can also 
account for the current data. In this paper we consider
the observational constraints on models for which in 
addition to the adiabatic mode either one, two, or three 
isocurvature modes are allowed as well as non-vanishing 
cross-correlations between the allowed modes. We consider 
only the isocurvature modes
present if no new physics is assumed---that is, the 
baryon isocurvature, the CDM isocurvature (CI), the neutrino 
isocurvature density
(NID), and neutrino isocurvature velocity (NIV) modes. The baryon and 
CDM isocurvature modes are not considered separately, because they imprint
virtually indistinguishable perturbations on the CMB. 
Consequently, without loss
of generality, we consider only the CI, NID and NIV modes. 

A simple family of models with only adiabatic perturbations
is defined by the cosmological parameters $\Omega _\Lambda,$  
$\Omega _b,$  $\Omega _{cdm}$ (subject to the constraint 
$\Omega _\Lambda +\Omega _b+\Omega _{cdm}=1),$ 
$h,$ and the reionization optical depth $\tau .$
These models may be compared to the WMAP data
\cite{wmapdata,wmapwebsite}, extended to smaller angular
scales by including data from the ACBAR \cite{acbar}, 
BOOMERANG \cite{boom}, CBI \cite{cbi} and VSA \cite{vsa}
experiments and LSS data from the 2dF galaxy redshift survey (2dFGRS), 
\cite{2dF_percival}.
The analysis by the WMAP team indicates a good fit to a model
of this sort with the following best-fit parameter choices:
$\Omega _\Lambda=0.74, \Omega _b=0.043, h=0.73, 
\tau=0.15$ \cite{spergeletal}. 

The object of this paper is to investigate models where the 
requirement of {\it adiabaticity} has been relaxed. {\it Adiabatic}
perturbations result when the stress-energy filling the 
universe obeys a single, spatially-uniform equation of state.
Initially, on surfaces of constant cosmic density, the densities
of each of the components are also uniform and share a common
velocity field. This is the simplest, but by no means only possible,
assumption about the character of the primordial perturbations.
The simplest, single-field models of inflation yield perturbations
that are exclusively adiabatic, but more complicated perturbations
are possible from multi-field inflationary models. Perturbations
in the initial equation of state or in the relative velocity
between the various components are called `isocurvature' because
asymptotically, at vanishing cosmic time, there is no perturbation
in the spacetime curvature, or more precisely this perturbation 
is of higher order. However, as the universe evolves, 
the various
components contributing to the stress-energy evolve differently,
leading to perturbations in the density, which source perturbations
in the gravitational potential, whose potential wells attract all
the components, leading to a gravitational instability. One might
say that the distinction between so-called {\it adiabatic} and 
{\it isocurvature} perturbations is useful only asymptotically
at very early times, and that almost any perturbation 
eventually 
leads to curvature perturbations. 

Since the CMB is imprinted at a rather late time in our early cosmic
history, at $z\gtorder 10^5,$ for comparing to the data, 
it suffices to consider a universe containing only baryons, CDM, photons, 
neutrinos and a cosmological constant $\Lambda. $ This seems to be the 
minimal physics at the relevant energy scales required to 
account for the present observations. There are many possibilities
for how the perturbations in these components may have been 
imprinted at early times by physics at higher energy scales
having other relevant degrees of freedom, but for our purpose,
as long as there are no relic particles (other than the CDM) 
left over from such an earlier epoch, the predictions for the 
CMB and LSS may be derived entirely from the perturbations
in the baryon, CDM, photon, and neutrino components. 
If modes that diverge at early times are excluded, there are
only five allowed modes: an adiabatic mode (AD), a baryon 
isocurvature mode (BI), a CDM isocurvature mode (CI) 
\cite{peebles,bond}, 
a neutrino density isocurvature mode (NID) and a 
neutrino velocity isocurvature mode (NIV) 
\cite{rebhan_schwarz_and_lasenby_challinor,bmt}. 
(For a review and further references see ref. \cite{bmt}). 

As long as only quadratic observables are considered, the 
perturbations in the five modes just enumerated may be 
completely characterized by a matrix valued power spectrum
\cite{langlois,bmt}
\ba
\biggl< X_I({\bf k})~X_J^*({\bf k}')\biggr> =
{\cal A}_{IJ}(\vert {\bf k}\vert )~\delta ^2({\bf k}-{\bf k}')
\ea
where the indices $I,J=1,2,3,4,5$ label the modes
AD, BI, CI, NID, NIV, respectively, 
and the random variable $X_I({\bf k})$ labels the amplitude of the 
$I$th mode with wavenumber ${\bf k}.$ Because balancing perturbations
in the CDM against the baryon density leads to a mode that hardly
evolves at all, we do not consider the baryon and CDM isocurvature
separately and opt for considering only the CDM isocurvature mode.
The diagonal elements represent auto-correlations whereas the 
off-diagonal elements represent cross-correlations. These must
on physical grounds be constrained so that $A_{IJ}$ is always 
positive definite. 

A variety of models have been proposed that generate mixtures 
of adiabatic and isocurvature perturbations with non-trivial
correlations. In an inflationary context, any
field of mass small compared to the Hubble scale during inflation
is disordered in a calculable manner, with significant power on
very large scales. This mechanism is exploited in the curvaton scenarios, 
where such a field is used to set up a variety of initial conditions 
\cite{motak,lyth}.
Another interesting possibility arises if inflation is driven
by two scalar fields \cite{gbwa,lmuk,langlois}. Recently, it was 
shown that
perturbations in the (dominant) scalar field which lie along the
direction of its motion correspond to adiabatic perturbations, while
orthogonal perturbations correspond to isocurvature perturbations
\cite{gwbm}. An arbitrary trajectory, in general, generates
correlated adiabatic and isocurvature perturbations with the magnitude
of the auto- and cross-correlation amplitudes depending on
the particular model. Other possibilities for generating isocurvature 
perturbations are discussed in \cite{bmt}.

It is interesting to ask whether such models may be distinguished
observationally. We adopt a model-independent approach,
attempting to constrain the amplitude of correlated
adiabatic and isocurvature perturbations directly from the data.
A previous investigation \cite{bmt_2002} estimated the quality of the 
constraints on isocurvature modes and their correlations 
that would be obtained from 
the WMAP and PLANCK data based on the published projected sensitivities
of these instruments. It was shown that when the adiabatic assumption
was relaxed, the uncertainties on the determination of the cosmological
parameters degraded substantially and large admixtures of isocurvature
were allowed.  Additional polarization data, however, was shown to
break the degeneracies introduced by isocurvature
modes, thereby constraining non-adiabatic amplitudes to less than
$10\%$ and recovering accurate parameter estimates. In this analysis
a fiducial model was assumed and estimates on the accuracy of
parameter estimation were obtained by examining the second partial
derivatives of the log likelihood about the most likely model.
Here we fully explore the entire parameter space using Markov Chain 
Monte Carlo (MCMC) methods
with the WMAP data, smaller scale CMB datasets and LSS data from 2dFGRS.

Observational constraints on isocurvature perturbations from
cosmological data have been studied previously. Earlier work
considered the viability of uncorrelated admixtures of adiabatic and
isocurvature modes \cite{lit_iso_preWMAP}. Prior to WMAP, limits
on correlated mixtures of the full set of adiabatic and isocurvature
modes were investigated, but over a smaller set of cosmological
parameters \cite{trotta_etal}. Recently, constraints on models in which
the adiabatic mode is correlated with a single isocurvature mode, have been
obtained using WMAP data \cite{lit_iso_postWMAP,crotty}. 

In this paper we undertake a full analysis within a common framework
of all combinations of the four allowed modes. 
We investigate the constraints on correlated isocurvature modes
that arise from these datasets and study the effect that isocurvature
modes have on parameter extraction. We consider models in which the
adiabatic and isocurvature modes share a common spectral
index. Although there exist more general classes of models that
predict independent spectral indices for adiabatic and isocurvature
modes, the reason for this restriction is that current data, as we
shall see, only weakly constrains the most general primordial
perturbation with a {\it common} spectral index for pure
modes. Allowing independent spectral indices for each pure mode would
further degrade parameter extraction due to complex flat directions
that would arise. Independent spectral indices were considered in
\cite{crotty} but only for a single correlated isocurvature mode. We
have reserved this extension for a future study, when such an analysis
will become more tractable as the quality of data improves. 

In considering the most general primordial perturbation we can
address one of our main concerns: whether it is possible to set
stringent bounds on the total isocurvature contribution to current
cosmological data. When considering specific models at the origin
of hybrid initial conditions, other
groups have tended to restrict themselves to at most one isocurvature
mode and its correlations with the adiabatic mode 
\cite{motak,lyth,gbwa,lmuk}. In general they have found that it is 
unlikely that
a considerable fraction of isocurvature perturbations is allowed. In this 
paper we present a comprehensive study of combinations of adiabatic and
isocurvature modes. We find that we essentially reproduce previous constraints 
if we only consider a reduced parameter space
with one adiabatic and one isocurvature mode. If we incrementally increase 
the number of modes, we find that the constraint on the fractional
contribution of isocurvature modes relaxes. Hence in this paper we
are able to place bounds on a wide range of theoretical models, 
complementing and extending the work presented in \cite{bdfms}.

The paper is organized as follows. Section \ref{Param} details
the parameterization and prior probability distribution
used for our analysis. The results are presented in section 
\ref{Results} for the various possible combinations of primordial 
perturbation modes. In section \ref{Priors} the sensitivity of
the results to various aspects of the prior probability 
distribution is investigated. Section \ref{Discussion} presents some 
general conclusions. 

The analysis in this paper required certain technical advances
in order to characterize the posterior probability distribution
in a computationally efficient and feasible manner. Because of 
the increase in 
the number of undetermined parameters compared to the case of
pure adiabatic models and the presence of poorly determined 
degenerate directions, it was not possible to carry out this 
analysis with the same technology used for MCMC chains for
the pure adiabatic models. The techniques used to tune the 
MCMC chains and diagnose their convergence have already
been described in a separate publication \cite{jo}. In  
an Appendix we detail the modifications made
to the publicly available DASh software \cite{DASh,DASh_web} used for calculating 
theoretical model spectra when isocurvature modes and their
correlations are admitted. With this software, 
cosmological models are first evaluated on a grid in parameter
space. Subsequently, cosmological models are evaluated by
interpolation over this grid.  

\section{Parameter Space and Datasets}
\label{Param}

\subsection{Cosmological parameters}

We assume four cosmological fluids: electromagnetic
radiation $\rho_\gamma$, baryons $\rho_b$, non-relativistic cold
dark matter $\rho_c$, and massless neutrinos $\rho_\nu,$ all
contributing to the total energy density.
For the unperturbed model, we define 
$\omega_i=\Omega_i\,h^2$ and
$\Omega_i=8\pi G\rho_i /3 H_0^2$
where $i$ labels the components, $\rho_i$ indicates the energy densities, 
and $H_0=100\, h\,\mbox{km\,s}^{-1}\,\mbox{Mpc}^{-1}$ is the present Hubble
constant. A contribution from a cosmological
constant $\Omega_\Lambda$ is also included. We restrict
our analysis to flat universes with $h>0.5.$
We consider the following parameter ranges:
$0.03<\omega_c<0.3$,\, $0.01<\omega_b<0.057,$ and
$0<\Omega_\Lambda<0.9.$ 
There is some evidence from a comparison of adiabatic models with the
WMAP data that an unconventional ionization history is preferred 
\cite{spergeletal}. In
particular a large optical depth $\tau \approx 0.17$ seems to indicate an
epoch of reionization which may be much earlier than previously
supposed. We therefore allow a wide range of admissible
optical depth, $0<\tau<0.3.$

\subsection{Mode parameterization}

We characterize the adiabatic and isocurvature perturbations using
the covariance matrix
\ba
\langle{ X}_I({\bf k})\,{ X}_{J}^*({\bf k'})\rangle
={\cal A}_{IJ}(k)\delta^3({\bf k}-{\bf k'}),
\ea
which provides a complete description of the primordial perturbations
for Gaussian initial conditions \cite{bmt}. 
Here $I$ may take the values AD, CI, NID, NIV, or some subset thereof 
containing 
$N$ pure modes. As previously discussed, the baryon
isocurvature mode need not be considered separately, so we do not include it. 
We assume that the
underlying power spectra are smooth over a wide range of $k$, so that
the diagonal and off-diagonal elements of the matrix ${\cal A}$ can be
parameterized as
\ba
{\cal A}_{IJ}(k) = A_{IJ}~k^{n_{IJ}}, \qquad n_{IJ} = \left\{ 
\begin{array}{cc}
n_I\, , & I=J, \\
{1 \over 2}(n_I+n_J)\, , & I \neq J,
\end{array} \right. 
\ea
in terms of a set of amplitude parameters $A_{IJ}$ and power-law 
shape parameters $n_I$. The cross-correlation
spectral indices are set to the mean of the corresponding auto-correlation spectral
indices to ensure that ${\cal A}$ remains positive semi-definite over a wide range of
scales. Our convention for the shape parameter is that scale-invariant 
adiabatic and isocurvature spectra correspond to $n_I=1.$
In practice this is achieved by choosing the random variable, $X_I$, that 
defines the initial power spectrum as $X_{AD}=\delta_\gamma, 
X_{CI}=k^2\delta_c,X_{NID}=k^2\delta_\nu,$ and $X_{NIV}=k^2v_\nu$ for the
different modes, where $\delta_\gamma, \delta_c$ and $\delta_\nu$ are the 
primordial photon, CDM and neutrino density contrasts, respectively, 
and $v_\nu$ is the neutrino velocity perturbation, at early times.
We assume a constant spectral index $n_S$ for all the modes, so that
\ba
{\cal A}_{IJ}(k) = A_{IJ}~k^{n_s}.
\ea
We parameterize so that 
\ba 
A_{IJ} \propto z_{IJ} 
\label{eqn:amplitude}
\ea
where
\ba {\rm tr}(zz^T)=\sum_{I,J=1}^N z_{IJ}^2=1
\ea 
and the proportionality constant that specifies the overall amplitude 
is defined below. The matrix elements
$z_{IJ}$ indicate the fractional contributions of the various auto- and
cross-correlations.
There are $D=N(N+1)/2$ independent correlations and
the coefficients $z_{IJ}$ cover a unit sphere of dimension $d=D-1$ on 
which we use the measure corresponding to the usual volume element.

For the MCMC random walk the coordinates 
$z_{IJ}$ are not useful. We therefore map the surface of the 
$d$-dimensional sphere $S^d$ 
onto a $d$-dimensional ball $B^d$ using the volume preserving mapping 
\ba
r(\theta )=\left[ d\int _0^\theta d\bar \theta ~\sin ^{(d-1)}\bar \theta \right] ^{1/d},
\ea
where $\theta $ is the angular coordinate with respect to the north pole of 
$S^d,$ the point corresponding to a pure adiabatic model and $r$
is the radial coordinate of $B^d$. Consequently, a random walk in 
$B^d$ (subject to the constraint $\theta <\pi$) may be lifted to 
$S^d$ by the inverse mapping. More explicitly,
\ba
z_{11}=\cos \theta, 
~z_{12}={\sin \theta \over \sqrt{2}}~{w_1\over r},
\ldots, 
~z_{1N}&=&{\sin \theta \over \sqrt{2}}~{w_{N-1}\over r},  
\cr
~z_{22}=\sin \theta ~{w_{N}\over r},
\ldots, 
~z_{2N}&=&{\sin \theta \over \sqrt{2}}~{w_{2N-2}\over r},
\cr
&& \vdots  
\cr
~z_{NN}&=&\sin \theta ~{w_{d}\over r}, 
\ea
and $z_{IJ} = z_{JI}$, where $w_i$ are the Cartesian coordinates 
of the Euclidean space into which
$B^d$ had been embedded, $r=\vert {\bf w}\vert,$ and $\theta =\theta (w).$ 
In practice, we choose $I=1$ to label the adiabatic mode, 
with $2 \dots N$ labeling 
the isocurvature modes. 
This parameterization covers the space of all symmetric 
$N\times N$ matrices.
We implement positive definiteness by assigning a zero prior
probability to any proposed matrix $z$ having a negative eigenvalue.
Since the eigenvalues can be computed quickly, any proposed steps
in the MCMC sequence are efficiently rejected before the corresponding
cosmological model is calculated. 

The modes are first normalized to give
equal power to the CMB anisotropy summed
from $\ell =2$ through $\ell =2000$ inclusive.
The power of an auto-correlated
mode $I$ is defined as
\ba
\sum _{\ell =2}^{2000}(2\ell +1)~C^{II}_\ell.
\ea 
For the cross-correlations
the geometric means of the respective renormalization factors is used. 
This defines the 
fractional contributions $z_{IJ}$ in terms of the physically observable 
power in each mode.

We now define the constant of proportionality in 
eqn.~\ref{eqn:amplitude}. We choose to sample the total power
in the CMB anisotropy, $A = \sum _\ell(2\ell +1)~C_\ell,$ where 
$C_\ell= \sum_{IJ}A_{IJ} C_\ell^{IJ},$
since it is one of the best determined quantities 
from the CMB data and hardly 
varies among models lying in the high-likelihood region of parameter space.
To do so we perform a second renormalization so that 
\ba
A_{IJ}= A \left[\sum_{\ell}\sum_{I,J}  (2\ell +1)~ z_{IJ}~C^{IJ}_\ell \right]^{-1}
~z_{IJ}~.
\ea

\begin{figure}[t]
\epsfig{file=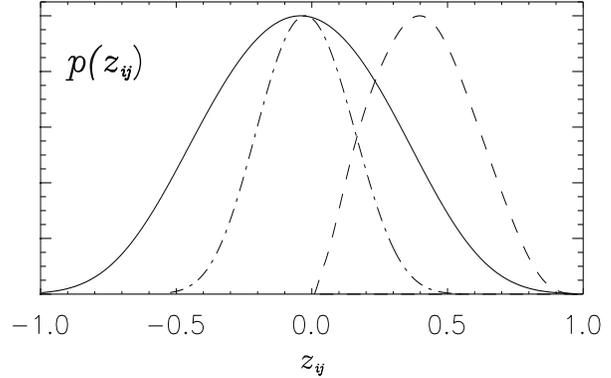,width=9cm,height=6cm}
\caption{
{\bf Shape of prior distribution for the observables 
$z_{\mdcs{I}{I}}$ and $z_{\mdcs{I}{J}}.$}
Our prior distribution of matrices subject to the constraint that 
$tr(zz^T)=1$ and that $z_{\mdcs{I}{J}}$ have only positive 
eigenvalues, in the absence of data, would give these posterior distributions
for the fractional auto-correlations $z_{\mdcs{I}{I}}$ (dashed) 
and cross-correlations $z_{\mdcs{I}{J}}$ (dot-dashed) for $N=4$.
The solid curve shows the distribution 
for $z_{\mdcs{I}{I}}$ that would result if the positive definiteness
requirement were suppressed.
\label{nodata_spd}}
\end{figure}

Except for the second renormalization, the prior distribution of $Z$  
is invariant under the transformations 
\ba 
Z\to OZO^{-1} 
\ea 
where $O\in SO(N).$
Other parameterizations may be contemplated. For example,
$Z$ may be expressed in the form
\ba
Z=ODO^{-1}
\ea
where $O$ is orthogonal and $D$ is diagonal. This is essentially
the approach proposed in \cite{trotta_etal}.\footnote{Actually,
in \cite{trotta_etal} a Cartesian measure is chosen for $\alpha _i$ where
$O=\exp [\alpha _i\Sigma _i].$ This measure, however, is not invariant
under group translations.}
The space of orthogonal matrices has a natural volume element, because
the requirement that volume be invariant under left (or right) translation
fixes the form of the volume element up to a constant factor. It is,
however, less obvious what measure should be chosen for the space
of diagonal matrices.

In our approach the effective measure for $D$ is
proportional to  \ba \delta \left( \sum _I\lambda
_I^2-1\right) ~~ \prod _{I<J} \vert \lambda _I-\lambda _J\vert,
\label{eqn:eval} \ea \cite{mehta} where $\lambda_I$ 
labels an eigenvalue of $D$. The second factor in the above 
expression gives rise to the well know phenomenon of
``level repulsion." Eigenvalues of a random matrix
do not follow Poisson statistics but rather are anti-bunched.
This level repulsion greatly
disfavors models for which one eigenvalue dominates, 
because in this case all the others are pushed toward zero. 
In section IV we shall consider rescalings
to counteract this phenomenon. 
The effect of this level repulsion is illustrated in Fig.~\ref{nodata_spd}
where the prior distributions for 
$z_{IJ}$ are plotted for $N=4.$ The solid line 
corresponds to $p(z_{IJ})$ for all modes if no positive definite 
requirement is placed on the matrix. For smaller $N$ the effect is less 
extreme. 
A further phase-space effect arises from the positive semi-definite
requirement, which constrains $Z$ to have positive eigenvalues.
This results in  a phase space suppression at low 
auto-correlation amplitudes $z_{II}$ and high cross-correlation 
amplitudes $z_{IJ}, I \ne J.$ The 
resulting prior distributions for these modes are shown in 
Fig.~\ref{nodata_spd}. 

Using these conventions we define the non-adiabatic
fraction as 
\ba
\fiso={ \ziso \over \ziso +z_{\mdcs{AD}{AD}}}
\ea
with the
isocurvature contribution to the data given by
\ba
\ziso=\sqrt{1-z_{\mdcs{AD}{AD}}^2},
\ea
which includes the contribution from cross-correlation modes. 

Another possibility, in terms of amplitudes, is to consider 
$\sqrt{\fiso}$ as an alternative candidate 
for the isocurvature fraction.
We note that other measures of the
non-adiabatic fraction are possible as well,
particularly in the presence of several modes
and their correlations.

Because of the possibility of cross-correlations, the total
power is not simply the sum of the adiabatic and the
isocurvature powers.
It is the amplitudes that add, the power being the square of
their sums. Consequently, constructive or destructive
interference or a combination of these may take place.
It is therefore theoretically possible to have virtually
identical adiabatic and total power in the presence
of a large isocurvature contribution.

\subsection{Datasets}

For the CMB data, we use the WMAP first-year
$C^{T}_\ell$ and $C^{TE}_\ell$ data \cite{wmapdata}
covering
$2\le \ell \le 800$ using the likelihood function in
\cite{verde} and
extend the data for 
$C^{T}_\ell$ to $\ell>800$ from the ground-based and balloon-borne
experiments ACBAR \cite{acbar}, BOOMERANG \cite{boom}, CBI \cite{cbi}, 
and VSA \cite{vsa},
using the compilation in
\cite{tegmarkcomp,tegmarkwebsite} 
with the covariance matrix for $\ell \le 800$ excised.

For the large-scale structure data, we use
the galaxy power spectrum ${\cal P}_g(k)$ as
measured in redshift space by the 2dFGRS experiment 
\cite{2dF_percival} at an effective survey redshift 
of $z=0.17$. 
We use only data in the linear regime $0.01 <
k/(h \,\mbox{Mpc}^{-1}) < 0.15$
and assume that the distortion 
due to redshift distortion 
and the bias are independent of scale. Following
\cite{verde} we rescale according to
\be
{\cal P}_g(k)|_{z=0.17}=f(b,\beta)~{\cal P}_m(k)|_{z=0}
\ee
where the rescaling factor
\be
f(b,\beta) =b^2(1+ \frac{2}{3}\beta_{eff} + \frac{1}{5}\beta_{eff} ^2)
\ee
is given in terms of the redshift distortion factor $\beta $ and bias $b$.
We approximate the bias as $b\approx \Omega_m^{0.6}/\beta$ and
take $\beta_{eff}= 0.85 \,\beta$
as the correction to $\beta$ at redshift $z=0.17$. 
For the LSS data from 2dFGRS data $\beta$ is an additional
parameter with a Gaussian prior $0.43 \pm 0.077$ \cite{twodf} 
broadened by 10\% as in \cite{verde} to allow for 
possible error in $\beta_{eff}.$ 
\begin{figure}[t!]
\begin{center}
\epsfig{file=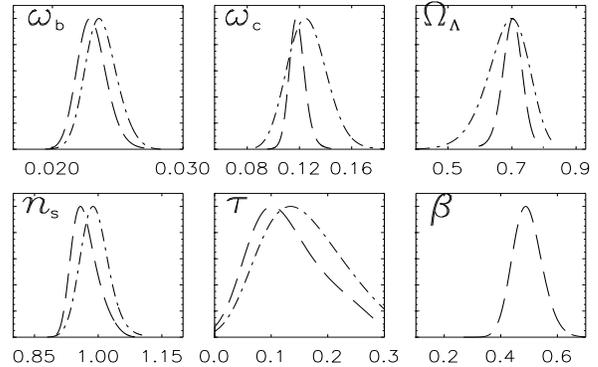,width=8cm,height=5cm}
\vskip +0.15in
\caption{{\bf Pure adiabatic model.}
Posterior distributions for the cosmological parameters 
using CMB data (dot-dashed) and CMB+LSS data (dashed).
\label{cosmo_hist_ad}}
\end{center}
\end{figure}

\begin{table}[t!]
\begin{center}

\begin{tabular}{lcc}
\hline
& \mbox{\fns CMB} & \mbox{\fns CMB+LSS} \\ \hline
$\omega_b$ & $0.024\pm 0.001$    & $0.023\pm 0.001$
            \\
$\omega_c$ & $0.13\pm 0.01$ & $0.120\pm 0.006$
            \\
$\Omega_\Lambda$ & $0.69^{+0.06}_{-0.08}$
& $0.71\pm 0.03$  \\
 $n_s$     & $0.99\pm0.03$  & $0.97\pm0.03$
           \\
$\tau$ &  $0.15\pm0.07$ & $0.13^{+0.08}_{-0.06}$
           \\
$\beta$ & $ \hld $  & $0.50\pm0.05$ \\
\\
$\Omega_m$ & {$0.31\pm^{ 0.08}_{0.06}$} & {$0.29\pm0.03$} \\
$h$ & {$0.69\pm0.05$} & {$0.70\pm 0.03$} \\
$b$ & $\hld$ & {$0.95\pm0.08$}  \\
\hline
\end{tabular}
\vskip 0.2in
\caption{{\bf Pure adiabatic model.} 
Median parameter values and $68\%$ confidence intervals 
using CMB data (column 1) and CMB+LSS data (column 2). 
\label{table_ad_only}}
\end{center}
\end{table}

\section{Results}
\label{Results}

\begin{figure}[t!]
\begin{center}
\epsfig{file=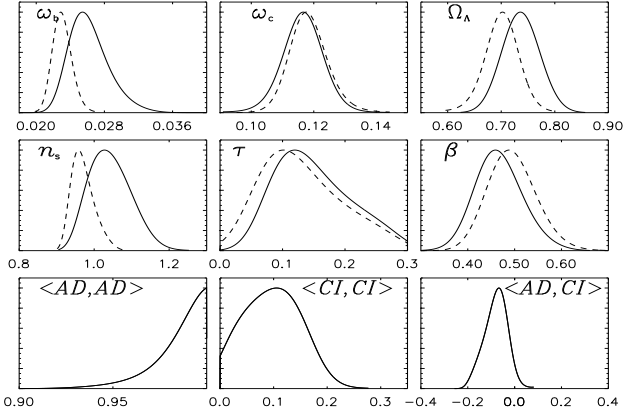,width=8cm,height=5.5cm}
\vskip +0.15in
\caption{
{\bf Adiabatic plus CI.}
We indicate the marginalized posterior distributions of the various parameters
for the adiabatic (AD) and CDM isocurvature (CI)
mixed models with correlations allowed (solid) using CMB+LSS,
with the result for the pure adiabatic model (dashed) included for
comparison.
$\mdcs{I}{J}$ denotes $z_{\mdcs{I}{J}}$.
\label{hist_ci}}
\end{center}
\end{figure}

In this section we present our results, first considering the pure
adiabatic model, essentially repeating the analysis of the WMAP
team, and then systematically introducing isocurvature modes
and their correlations. We first consider the addition of
a single isocurvature mode, then the addition of pairs of
isocurvature modes, and finally the addition of
all three allowed isocurvature modes. 

Fig.~\ref{cosmo_hist_ad} shows
the marginalized posterior probability 
distributions for the cosmological 
parameters, both for the CMB dataset
and for the combined CMB+LSS dataset, for pure adiabatic models. 
The median values and 68\%
confidence intervals for each parameter are given in
Table~\ref{table_ad_only}.
Our results are consistent with the WMAP
analysis \cite{spergeletal}.
The density parameter values $\omega _i$ inferred from CMB and LSS measurements
are consistent with those inferred using other techniques, such
as measurements of the baryon density from observations of primordial
light element abundances \cite{bbn} and of the energy density 
of a negative pressure
component inferred from Type Ia supernovae
\cite{sn_lambda}. Although the data suggests an earlier
reionization indicated by a large optical depth
$\tau,$ this parameter is not very well constrained by the CMB.
The tilt of the adiabatic power
spectrum is also well measured, with a nearly scale-invariant spectrum
favoured.
Table~\ref{table_ad_only} 
also lists several derived parameters 
such as the matter fraction $\Omega_m$, 
the dimensionless Hubble parameter $h$, 
and the bias $b.$

\subsection{Adiabatic mode correlated with single isocurvature modes}
\label{one_mode_section}

We present results for each of the cold dark matter isocurvature (CI),
neutrino isocurvature density (NID) and neutrino isocurvature velocity
(NIV) modes correlated with the adiabatic mode. The median values and
the 68\% confidence intervals of the cosmological parameters and mode
amplitudes are given in Table~\ref{table_ad_1iso} for each of the
above mode combinations. Of immediate interest are the relative powers
that are permitted for the pure isocurvature modes and their
cross-correlations. In all three cases the preferred models are
dominated by the adiabatic mode, with a small non-adiabatic fraction,
$\fiso \sim 10\%$, tolerated.

\begin{table}[t!]
\begin{tabular}[h]{lccc}
\hline\hline
& \mbox{\fns AD+CI} & \mbox{\fns AD+NID}
&\mbox{\fns AD+NIV}\\ \hline
$\omega_b$ & $0.026\pm 0.003$ & $0.025\pm0.002$ &
     $0.026\pm 0.002$\\
$\omega_c$   & $0.116\pm 0.007$ & $0.118\pm0.007$
&   $ 0.112\pm 0.007$ \\
$\Omega_\Lambda$ & $0.74\pm 0.04$  & $0.73\pm 0.04$
& $0.71\pm 0.03$  \\
 $n_s$    & $1.04\pm 0.06$ & $1.02\pm 0.05$
& $ 1.02\pm 0.04$  \\
$\tau$  & $0.14^{+0.07}_{-0.05}$ &
$0.14^{+0.08}_{-0.06}$  &  $0.21 ^{+0.06}_{-0.09}$\\
$\beta$ & $0.46\pm 0.05$  & $0.48\pm 0.06$
&   $0.47\pm 0.05$  \\ \\
 $z_{\mdcs{AD}{AD}}$  & $0.99^{+ 0.009}_{- 0.02}$
&  $0.997^{+0.003}_{-0.006}$ & $0.99^{+0.008}_{-0.02}$ \\
$z_{\mdcs{CI}{CI}}$  & $0.10\pm 0.06$
& $\hld$ & $\hld$  \\
$z_{\mdcs{NID}{NID}}$ & $ \hld $
& $0.05\pm0.04$ & $\hld$ \\
$z_{\mdcs{NIV}{NIV}}$ & $ \hld $
& $\hld$ & $0.07^{+ 0.06}_{- 0.04}$  \\
$z_{\mdcs{AD}{CI}}$& $-0.08\pm0.05$
 &  $\hld$ &$\hld$\\
$z_{\mdcs{AD}{NID}}$  & $ \hld $
 & $-0.03\pm0.03$ &  $\hld$  \\
$z_{\mdcs{AD}{NIV}}$ & $ \hld $
& $\hld$ &  $0.08\pm0.08$ \\
\\
$z_{\mbox{\fnsc ISO}}$   & $0.15\pm 0.08$ & $0.08\pm 0.05$ & 
$0.15\pm^{0.10}_{0.07}$ 
 \\
$f_{\mbox{\fnsc ISO}}$  &  $0.13\pm 0.06$ &  
$0.07\pm 0.04$ & $0.13\pm 0.07$   \\
\\
$\Omega_m$  & $0.26\pm 0.04$ & $0.27\pm 0.04$ &
$0.29\pm 0.03$ \\
$h$   & $0.74\pm 0.05$ & $0.73\pm 0.05$ &
$0.69\pm 0.04$ \\
$b$ & $0.96\pm 0.08$  &  $0.94\pm 0.08$ &
$1.00\pm 0.10$  \\
\\
\hline
\end{tabular}
\vskip 0.1in
\caption{
{\bf Adiabatic plus one isocurvature mode.}
Median parameter values and $68\%$ confidence intervals
for mixed models with the adiabatic mode (AD) and a single 
isocurvature mode (columns 1-3), with
the CMB+LSS dataset used throughout. 
\label{table_ad_1iso}}
\end{table}

\begin{figure}[t!]
\begin{center}
\epsfig{file=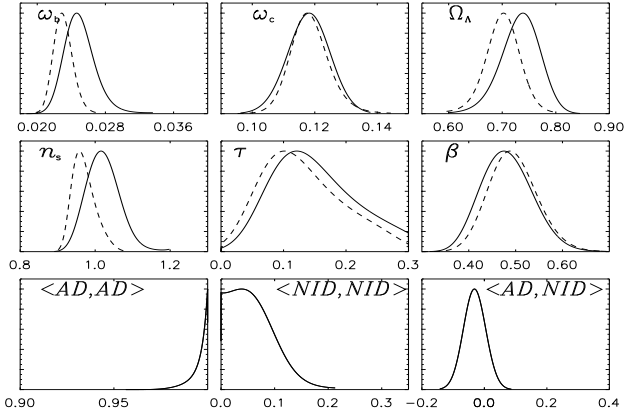,width=8cm,height=5.5cm}
\vskip +0.15in
\caption{
{\bf Adiabatic plus NID.} 
Distributions as described in Fig.~\ref{hist_ci}, for correlated 
adiabatic and neutrino density isocurvature models.
\label{hist_nid}}
\end{center}
\end{figure}

\begin{figure}
\begin{center}
\epsfig{file=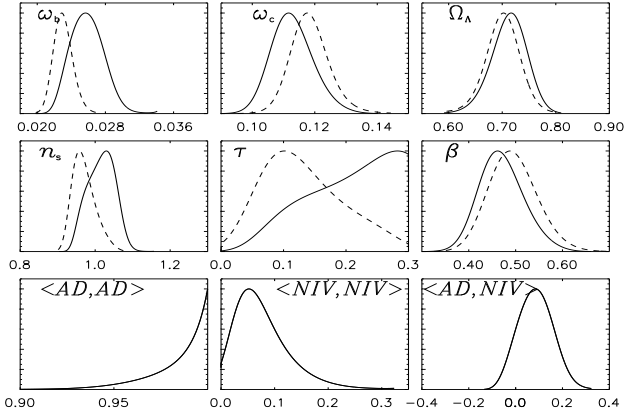,width=8cm,height=5.5cm}
\vskip +0.15in
\caption{{\bf Adiabatic plus NIV}.
Distributions as described in Fig.~\ref{hist_ci}, for correlated adiabatic 
and neutrino velocity isocurvature models.
\label{hist_niv}}
\end{center}
\end{figure}

The relative powers $z_{IJ}$ of the individual isocurvature
auto-correlation and cross-correlation modes are consistent with zero
at the 1$\sigma$ to 2$\sigma$ level. The marginalized parameter
distributions for the CI, NID and NIV modes are shown in 
Figs.~\ref{hist_ci}, \ref{hist_nid}, and \ref{hist_niv}, respectively. The
best fit mixed models (which have a 10\% non-adiabatic contribution) do
not fit the data significantly better than the best-fit adiabatic
models. An illustration of the flat direction that arises in
correlated adiabatic and CI models was given in \cite{gwbm}.

Including a single correlated isocurvature mode does not 
significantly change
the cosmological parameters from those in
pure adiabatic models, as shown in Table ~\ref{table_ad_1iso}.
The $n_s$ distribution is slightly broadened and shifted to 
larger values, and 
$\omega _b$ is slightly increased. 
For the NIV mode $\tau $ shifts toward larger values, and the distribution
appears cut-off by the limit at $\tau =0.3$ from the prior.
This is because the NIV mode adds most of its power on 
small scales ($\ltorder 1^\circ$) and its contribution to the 
CMB is damped by the larger $\tau .$ 

\begin{table}[t!]
\begin{tabular}{lccc}
\hline\hline
   & \mbox{\fns AD+CI+NID} & \mbox{\fns AD+CI+NIV} 
& \mbox{\fns AD+NID+NIV}\\
\hline    
$\omega_b$  & $0.030\pm 0.004$  & $0.032\pm 0.004$ &  $0.036\pm 0.005$ \\
$\omega_c$  & $0.128\pm^{0.011}_{0.009}$ &  $0.110\pm
0.007$ &  $0.115\pm
0.009$\\
$\Omega_\Lambda$  & $0.74\pm0.03$  & $0.73\pm0.03$ & $0.73\pm0.03$  \\
 $n_s$      & $1.00\pm0.07$  & $1.13\pm0.07$        &  $1.12\pm0.04$\\
$\tau$  & $0.16\pm0.07$  & $0.19\pm0.08$ &
$0.22\pm^{0.06}_{0.08}$  \\
$\beta$   &  $0.45\pm 0.05$  &  $0.40\pm 0.06$ & $0.38\pm 0.05$\\
\\
 $z_{\mdcs{AD}{AD}}$   &
$0.75\pm^{0.16}_{0.21}$ &  $0.89\pm^{0.07}_{0.13}$ &  $0.78\pm^{0.12}_{0.16}$\\
$z_{\mdcs{CI}{CI}}$ & $0.35\pm0.16$ & $0.14\pm0.09$ & $\hld$ \\
$z_{\mdcs{NID}{NID}}$  & $0.19\pm0.10$ & $\hld$ & $0.19\pm0.10$\\
$z_{\mdcs{NIV}{NIV}}$ & $\hld$ & $0.19\pm^{0.15}_{ 0.10}$ & $0.35\pm0.16$\\
$z_{\mdcs{AD}{CI}}$   & $-0.21\pm0.10$ &  $0.02\pm0.10$& $\hld$ \\
$z_{\mdcs{AD}{NID}}$  & $0.12\pm^{0.08}_{0.10}$& $\hld$ 
& $0.03\pm0.12$\\
$z_{\mdcs{AD}{NIV}}$  & $\hld$  & $0.24\pm0.13$  & $0.26\pm0.13$ \\
$z_{\mdcs{CI}{NID}}$  & $-0.22\pm0.11$ & $\hld$  &$\hld$ \\
$z_{\mdcs{CI}{NIV}}$  & $\hld$ & $0.03\pm0.05$  &$\hld$ \\
$z_{\mdcs{NID}{NIV}}$   & $\hld$ &  $\hld$ & $0.11\pm0.07$\\
\\
$z_{\mbox{\fnsc ISO}}$  & $0.66\pm^{0.18}_{ 0.26}$ & $0.46\pm^{0.19}_{ 0.16}$ &
$0.62\pm^{0.16}_{0.19}$\\
$f_{\mbox{\fnsc ISO}}$  & $0.47\pm0.16$ & $0.34\pm0.12$ & $0.44\pm0.12$\\
\\
$\Omega_m$ & $0.26\pm 0.03$ & $0.27\pm 0.03$&  $0.27\pm 0.03$ \\
$h$    & $0.78\pm0.06$ & $0.72\pm0.05$& $0.75\pm0.04$\\
$b$ & $1.00\pm0.10$ & $1.13\pm0.14$ & $1.21\pm0.16$ \\
\hline
\end{tabular}
\vskip 0.1in
\caption{{\bf Adiabatic plus two isocurvature modes.}
The same conventions as in Table \ref{table_ad_1iso} are used, for mixed 
models including two correlated isocurvature modes.
\label{table_ad_2iso}}
\end{table}

\begin{figure}[t!]
\begin{center}
\epsfig{file=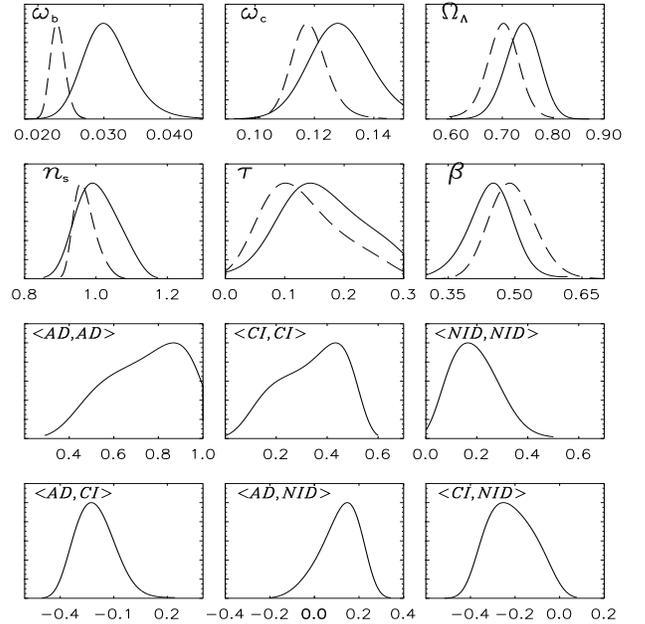,width=8cm,height=8.5cm}
\vskip +0.1in
\caption{
{\bf Adiabatic plus CI+NID.}
Distributions as described in Fig.~\ref{hist_ci} for the adiabatic mode 
correlated with both the CDM and neutrino density isocurvature modes.
\label{hist_ci_nid}}
\end{center}
\end{figure}

\begin{figure}
\begin{center}
\epsfig{file=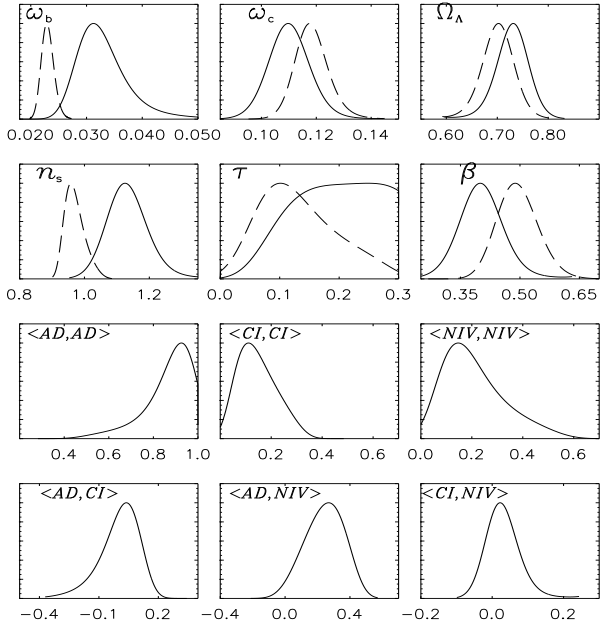,width=8cm,height=8.5cm}
\vskip +0.1in
\caption{
{\bf Adiabatic plus CI+NIV.}
Distributions as described in Fig.~\ref{hist_ci} for the adiabatic mode 
correlated with both the CDM density and neutrino velocity isocurvature 
modes.
\label{hist_ci_niv}}
\end{center}
\end{figure}

\subsection{Adiabatic mode correlated with two isocurvature modes}
\label{two_mode_section}

\begin{figure}
\begin{center}
\epsfig{file=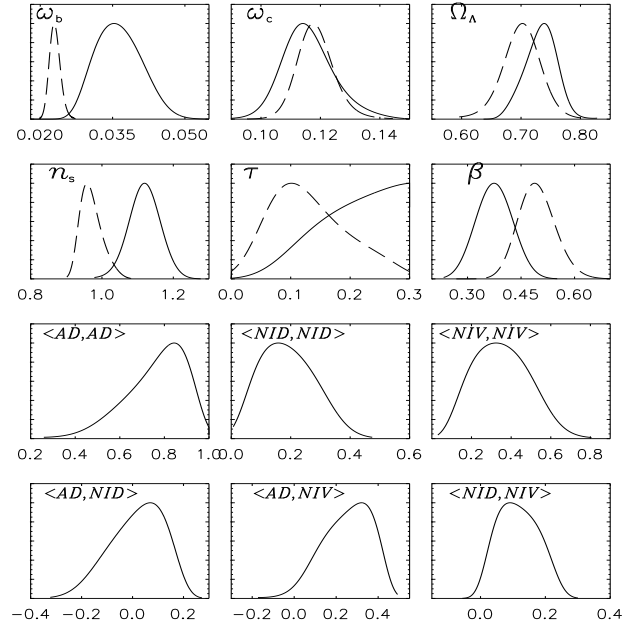,width=8cm,height=8.5cm}
\vskip +0.1in
\caption{
{\bf Adiabatic plus NID+NIV.}
Distributions as described in Fig.~\ref{hist_ci} for the adiabatic mode 
correlated with both the neutrino density and velocity isocurvature 
modes.
\label{hist_nid_niv}}
\end{center}
\end{figure}

We now study the effect of introducing two isocurvature modes
correlated with the adiabatic mode and each other.
The three possible combinations are CI+NID, CI+NIV, and NID+NIV.
In all cases the
isocurvature fraction increases significantly, to $\fiso \approx 
47\% , 34\%$ and $44\%$ respectively, 
as can be seen from Table~\ref{table_ad_2iso}. 
The posterior distributions are shown in 
Figs.~\ref{hist_ci_nid}, \ref{hist_ci_niv} and \ref{hist_nid_niv}.

\begin{figure}
\begin{center}
\epsfig{file=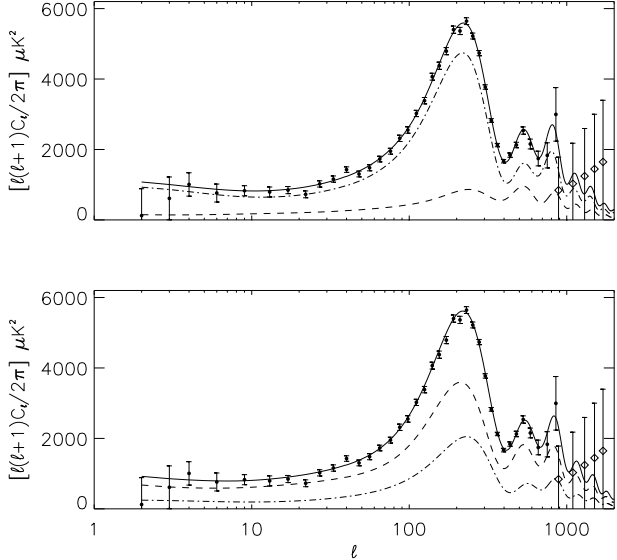,width=9cm,height=8cm}
\caption{
{\bf CMB temperature spectra for best fit mixed models.} (Top) 
The CMB angular power spectrum, $C^{TT}_\ell$, for 
the best-fit mixed model (solid)
including {\bf AD+CI+NID} modes with parameters ($\omega_b, \omega_d,
\Omega_\Lambda, n_s, \tau, \fiso$) = (0.033, 0.13, 0.75, 0.99,
0.16, 0.62). The adiabatic (dot-dashed) and 
non-adiabatic (dashed) contributions are included.  (Bottom) The same 
for models with {\bf AD+NID+NIV},
for parameters (0.037, 0.11, 0.74, 1.11, 0.22, 0.56). The CMB data is included.
\label{twomode_cl}}
\end{center}
\end{figure}

The isocurvature contribution is greatest for the 
CI+NID models, although apart from an 
increase in the baryon 
density, the cosmological parameters are not shifted 
significantly from their 
median values in the adiabatic case. Despite the large isocurvature 
fraction, the best-fit models have both positive and negative 
mode correlations which cancel to give a small non-adiabatic power, 
as illustrated in
Fig.~\ref{twomode_cl}, leaving the adiabatic mode with a power 
closely matched to the total spectrum. 
To gain insight into these non-adiabatic cancellations, we
utilise the covariance matrix of the distribution to perform a
principal component analysis about a high likelihood mixed model with
parameters ${\bar x}_i.$
We calculate the eigenvalues,
$\lambda_\alpha,$ and eigenvectors, ${\bf y}_\alpha,$ of the
covariance matrix,
\be
C_{ij}= \la \hat{x}_i \hat{x}_j \ra - \la \hat{x}_i \ra \la
\hat{x}_j \ra 
\ee
 where $\hat{x}_i=x_i/{\bar{x}_i}$ are
normalized by their fiducial values.  The flattest direction, ${\bf
y}_{0},$ along which changes in the CMB and matter power spectra are
negligible, has percentage error $|\delta {\bf y}_0| \equiv
{100\,\%/ \sqrt{\lambda_0} } =90\%.$ Fig.~\ref{degen_ci} shows the 
cancellation in the CMB temperature spectrum in this
direction, dominated by a 
subset of seven
parameters that comprises the spectral index, $n_s,$ and the power in
the six mode correlations. The increased 
auto-correlation power on large-scales that results from a low
spectral index is compensated almost exactly by the negative 
cross-correlation spectra
$\mdcs{AD}{CI}$ and $\mdcs{AD}{NID}$. This highlights the
important role played by these cross-correlations in allowing a large
isocurvature contribution.

The CI+NIV models allow a lower isocurvature fraction than those with 
CI+NID, but the baryon density, spectral index and optical depth $\tau$ 
are further increased and the redshift distortion parameter 
reduced from the pure adiabatic case. As observed in
the previous section, the increase in $\tau$ is correlated with an
increase in amplitude of the NIV mode. The increased NIV contribution
is also strongly correlated with an increase in the values of the
spectral index, $n_s,$ and the baryon fraction, $\omega_b$, which are
all associated with a flat direction in parameter space, as we will
show in the next section.

The cosmological parameters change more significantly from their
adiabatic values when both neutrino isocurvature modes are
introduced, shown in Fig.~\ref{hist_nid_niv}.  The baryon density,
spectral index and reionization parameter are significantly increased,
while the redshift distortion parameter is further reduced. 
The best-fit models have positive-valued cross-correlations, adding 
constructively as indicated in Fig.~\ref{twomode_cl} to produce 
a significant non-adiabatic amplitude.
The lower value of $\beta$ in these models
is then consistent with the reduced adiabatic power and higher
normalization factor $f(b,\beta)$ required to fit the LSS data.

\subsection{Adiabatic mode correlated with all three isocurvature modes}

\begin{figure}[t!]
\epsfig{file=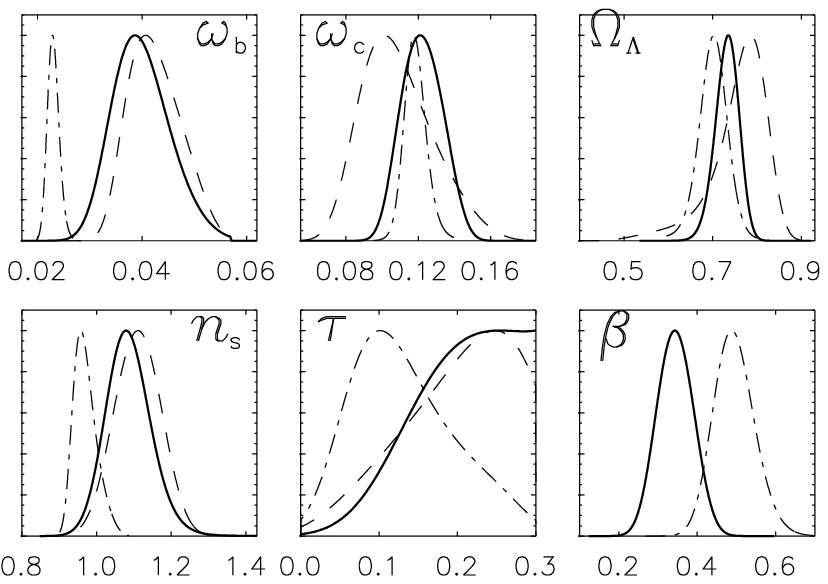,width=8cm,height=5cm}
\vskip +0.1in
\epsfig{file=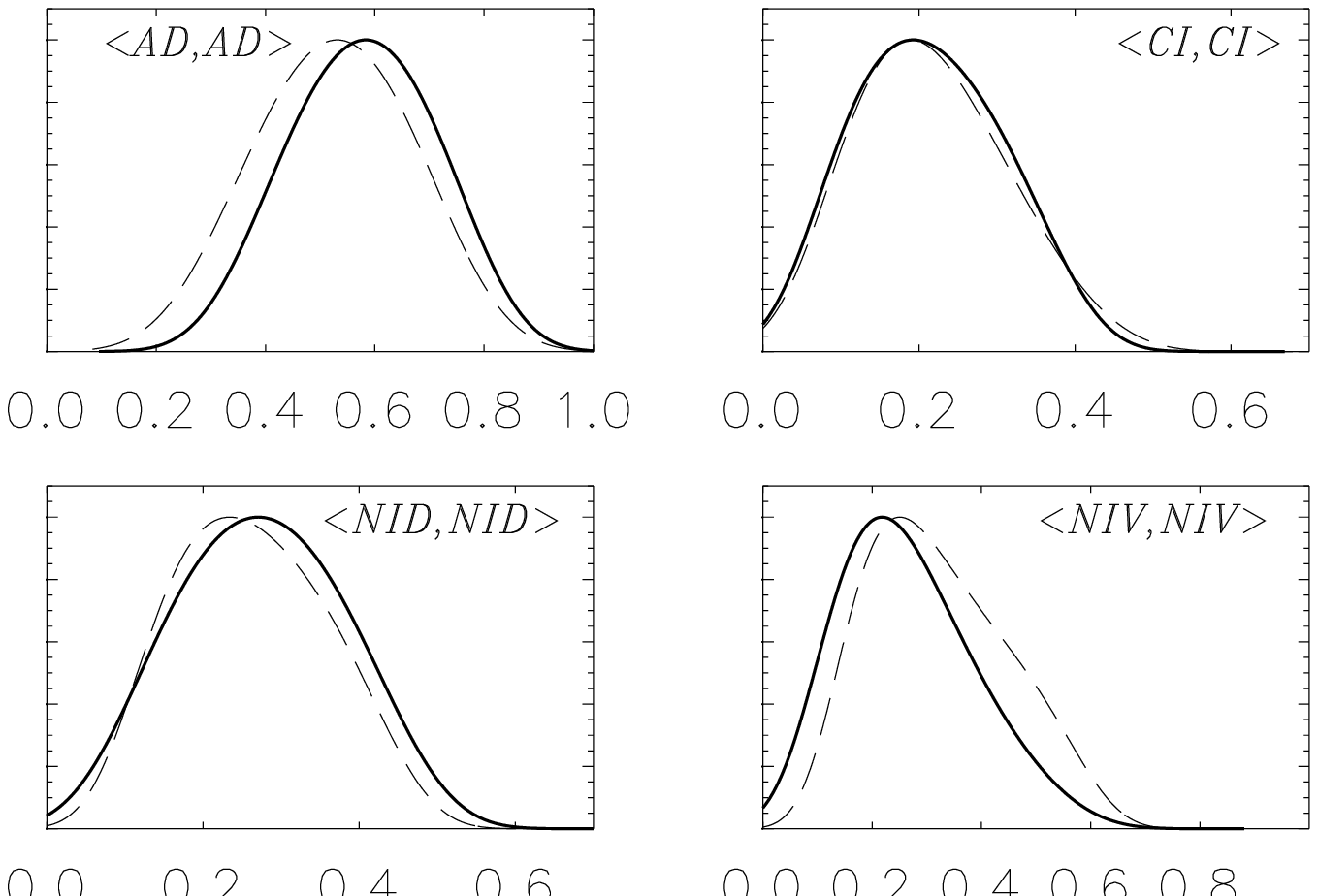,width=8cm,height=5cm}
\vskip +0.15in
\epsfig{file=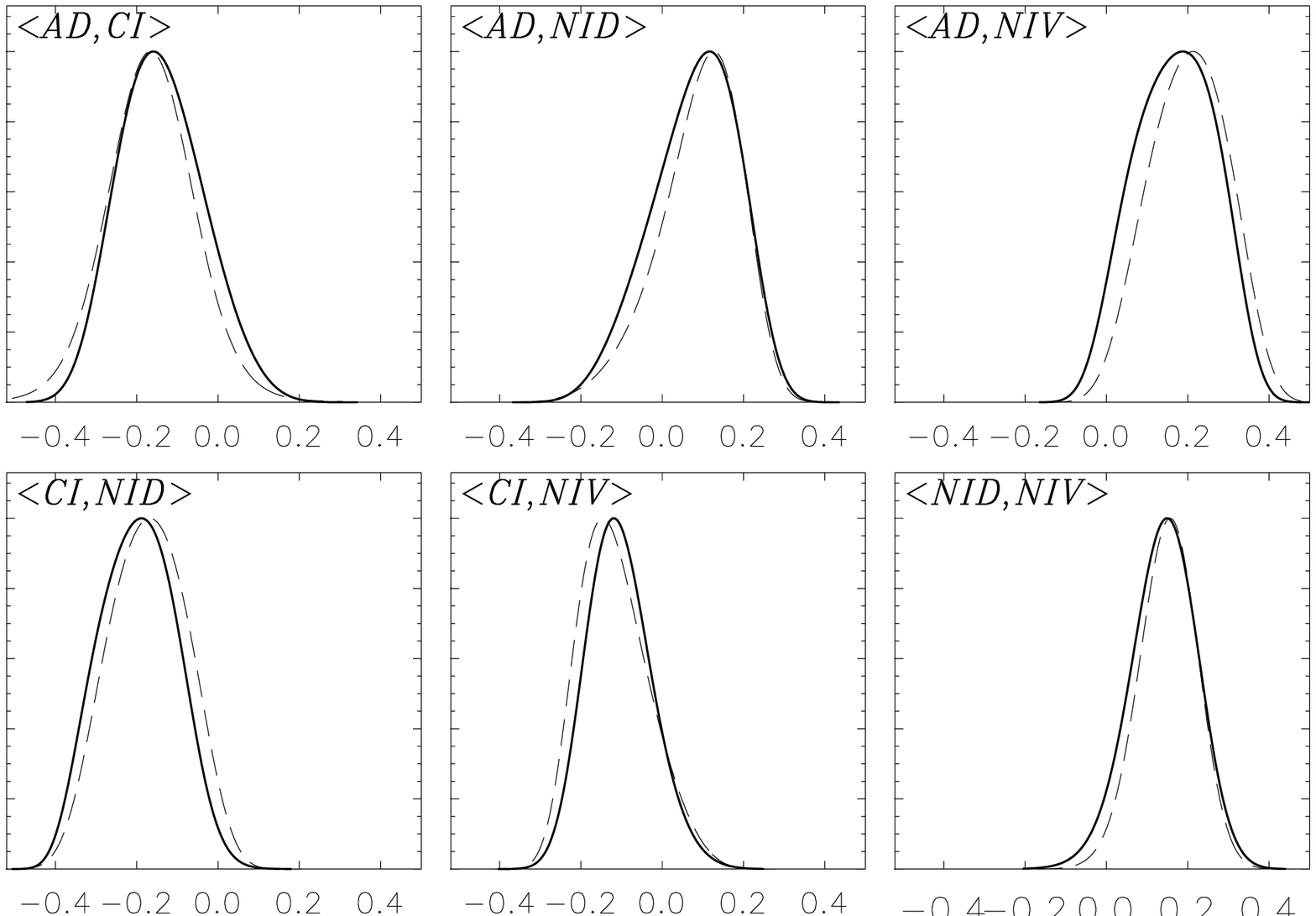,width=8cm,height=5cm}
\vskip +0.15in
\caption{
{\bf Posterior distributions for Adiabatic plus CI+NID+NIV.}
The distributions for models with correlated adiabatic and three 
isocurvature modes are shown, using the CMB only (dashed) and (CMB+LSS)
(solid). Where relevant the posterior for the
pure adiabatic models (dot-dashed) is shown for comparison.
\label{cosmo_hist}}
\end{figure}

We now consider a general perturbation comprising a linear combination
of the AD, CI, NID and NIV modes. Table~\ref{table_ad_3iso}
gives  statistics for the cosmological parameters and relative mode
amplitudes for both the CMB and combined CMB+LSS datasets, and these
distributions are plotted in 
Fig.~\ref{cosmo_hist}.

We observe that the adiabatic mode is no longer completely dominant.
The general model includes a mixture of adiabatic and isocurvature
modes with $\fiso = 60\%$ for CMB data alone which decreases
slightly to $\fiso= 57\%$ for the CMB+LSS dataset. 
The mixed model family  include models with
baryon fractions greater than that 
inferred from nucleosynthesis measurements as well as 
large optical depths and large
departures from scale-invariance. These departures
from the pure adiabatic parameter values are 
all correlated with larger
isocurvature fractions and have broad distributions. These shifts are
associated with a degenerate direction in parameter space which we
investigate below. The remaining cosmological parameters $\omega_c$
and $\Omega_\Lambda$ are consistent with their values in the
adiabatic case for the CMB+LSS dataset. This is also true for the
derived parameters, though there is a preference for slightly smaller
values of $\Omega_m$ and slightly larger values of $h$. 

\begin{figure}[t!]
\epsfig{file=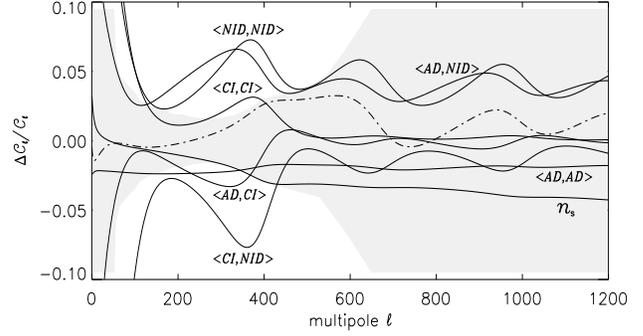,width=8.5cm,height=5cm}
\caption{
{\bf Derivative along flattest direction of the CI+NID  
distribution} 
(dot-dashed) is decomposed into variations in the 
mode contributions, ${\mdcs{I}{J}},$ and the overall spectral index $n_s$. 
The shaded region corresponds to the uncertainty of the WMAP
measurement including cosmic variance.
\label{degen_ci}} 
\end{figure}

\begin{table}[t!]
\begin{minipage}[h]{.48\textwidth}
\begin{center}
\begin{tabular}[h]{lccc}
\hline\hline
& \hspace{-4mm} \mbox{\fns AD} & \mbox{\fns AD+ISO} & \mbox{\fns
AD+ISO}\\
& \hspace{-4mm} \mbox{\fns CMB+LSS} & \mbox{\fns CMB}
& \mbox{\fns CMB+LSS} \\ \hline
\hspace{0.25cm}$\omega_b$ & \hspace{-4mm}
\mbs{$0.023\pm 0.001$}  &  \mbs{$ 0.043\pm 0.006$}  &
\mbs{$0.041\pm 0.006$} \\
\hspace{0.25cm}$\omega_c$ & \hspace{-4mm}
\mbs{$0.120\pm 0.006$} & \mbs{$0.11\pm0.02$} & \mbs{$0.12\pm0.01$} \\
\hspace{0.25cm}$\Omega_\Lambda$ & \hspace{-4mm}
\mbs{$0.71\pm 0.03$} & \mbs{$0.79\,$}$^{+\,0.05}_{-\,0.07}$  & \mbs{$0.74\pm0.03$} \\
 \hspace{0.25cm}$n_s$ & \hspace{-4mm}
\mbs{$0.97\pm0.03$} & \mbs{$1.13\pm0.07$}  & \mbs{$1.10\pm0.06$} \\
 \hspace{0.25cm}$\tau$ & \hspace{-4mm} \mbs{$0.13\,$}
$^{+\,0.08}_{-\,0.06}$ &
\mbs{$0.21\,$}$^{+\,0.06}_{-\,0.08}$ &  \mbs{$0.22\pm0.07$} \\
 \hspace{0.25cm}$\beta$ & \hspace{-5mm}
\mbs{$0.50\pm0.05$} & $\hld $ & \mbs{$0.35\pm0.05$} \\
\\
$z_{\mdcs{AD}{AD}}$ & \hspace{-4mm}
\hspace{1mm} \mbs{1.0} &
\mbs{$0.55\,$}$^{+\,0.16}_{-\,0.14}$ & \mbs{$0.61\pm{0.15}$} \\
$z_{\mdcs{CI}{CI}}$ & \hspace{-5mm} $ \hld $ &
\mbs{$0.23\,$}$^{+\,0.11}_{-\,0.09}$ & \mbs{$0.23\pm0.11$} \\
$z_{\mdcs{NID}{NID}}$ & \hspace{-5mm} $ \hld $ &
\mbs{$0.28\,$}$^{+\,0.12}_{-\,0.10}$ & \mbs{$0.30\pm0.12$} \\
$z_{\mdcs{NIV}{NIV}}$ & \hspace{-5mm} $ \hld $ &
\mbs{$0.34\pm0.14$} & \mbs{$0.28\,$}$^{+\,0.14}_{-\,0.11}$ \\
$z_{\mdcs{AD}{CI}}$ & \hspace{-5mm} $ \hld $ &
\mbs{$-0.14\pm0.10$} &  \mbs{$-0.12\,$}$^{+\,0.12}_{-\,0.10}$ \\
$z_{\mdcs{AD}{NID}}$  & \hspace{-5mm} $ \hld $ &
\mbs{$0.12\,$}$^{+\,0.09}_{-\,0.11}$ & \mbs{$0.11\,$}$^{+\,0.10}_{-\,0.12}$ \\
$z_{\mdcs{AD}{NIV}}$ & \hspace{-5mm} $ \hld $ &
\mbs{$0.22\pm0.10$} & \mbs{$0.19\pm0.11$} \\
$z_{\mdcs{CI}{NID}}$  & \hspace{-6mm} $ \hld $ &
\mbs{$-0.15\pm0.10$} & \mbs{$-0.18\pm0.10$} \\
$z_{\mdcs{CI}{NIV}}$ & \hspace{-6mm}  $ \hld $ &
\mbs{$-0.10\,$}$^{+\,0.10}_{-\,0.08}$ & \mbs{$-0.09\pm0.08$} \\
$z_{\mdcs{NID}{NIV}}$ & \hspace{-5mm} $ \hld $ &
\mbs{$0.17\pm0.07$} & \mbs{$0.16\pm0.08$} \\
\\
 $z_{\mbox{\fnsc ISO}}$  &$\hld$ &\mbs{$0.84\pm^{0.08}_{0.13}$} &
\mbs{$0.79\pm^{0.09}_{ 0.13}$}\\
$f_{\mbox{\fnsc ISO}}$ & $\hld$  &  \mbs{ $0.60\pm^{0.09}_{0.11}$} &
\mbs{$0.57\pm 0.09$} \\
\\
$\Omega_m$ &  \mbs{$0.29\pm0.03$} &
\mbs{$0.21\pm^{0.07}_{0.05}$} &  \mbs{$0.26\pm0.03$} \\
$h$ & \mbs{$0.70\pm 0.03$}  & \mbs{$0.85\pm^{0.06}_{0.08}$}
& \mbs{$0.80\pm 0.05$}\\
$b$ &  \mbs{$0.95\pm0.08$} &  $\hld$
& \mbs{$1.3\pm 0.2$}  \\
\hline
\end{tabular}
\end{center}
\end{minipage}
\vskip 0.1in
\caption{
{\bf Adiabatic plus all three isocurvature modes.} The same conventions
are used as in Table~\ref{table_ad_1iso}, for mixed models with all 
three correlated isocurvature modes (AD+ISO). Results are included for both 
CMB and CMB+LSS datasets, and 
for pure adiabatic models (AD) for comparison.
\label{table_ad_3iso}}
\vskip -0.2in
\end{table}

\begin{figure}[t!]
\begin{center}
\epsfig{file=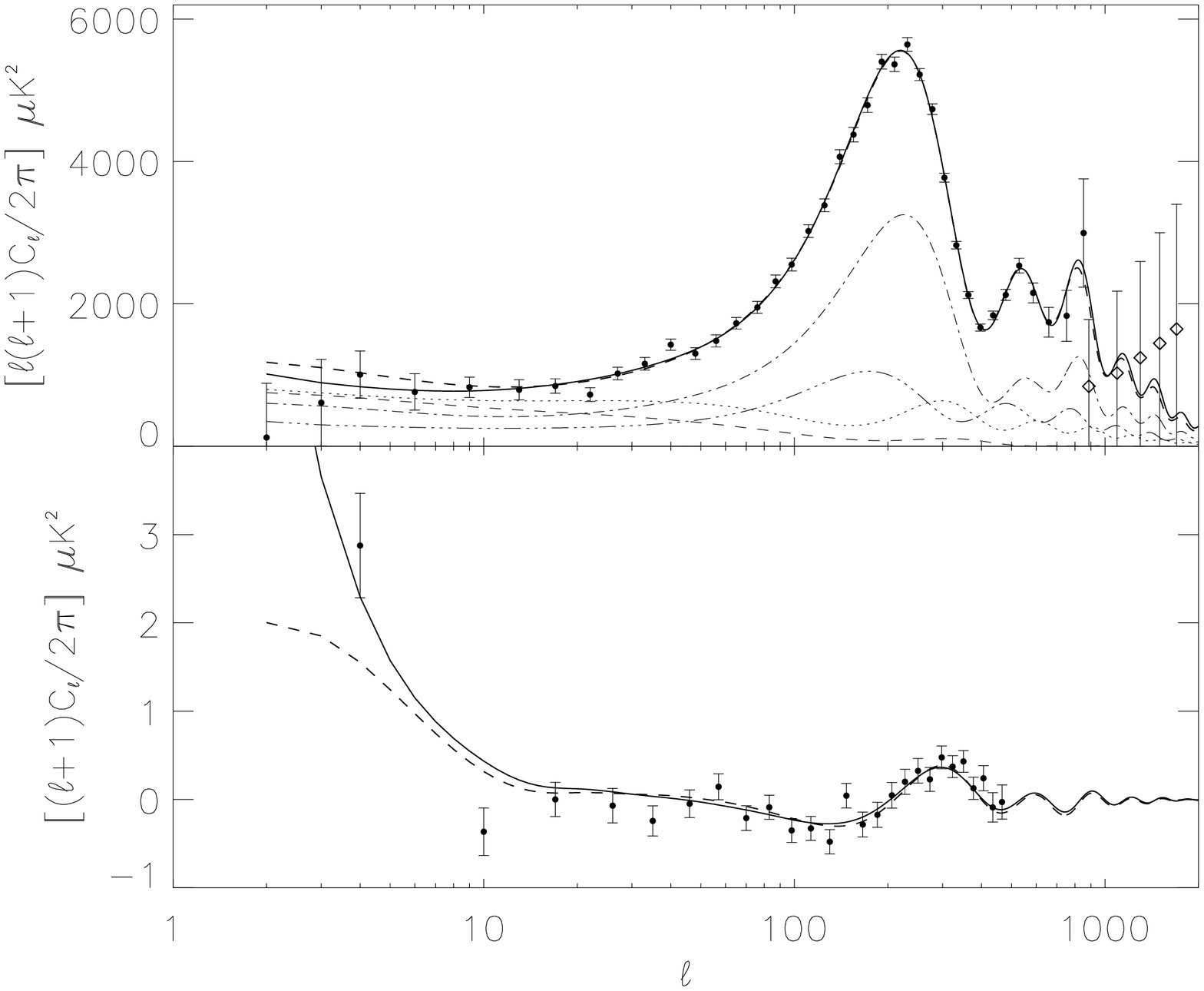,width=8.5cm,height=8.5cm}
\vskip -0.2in
\epsfig{file=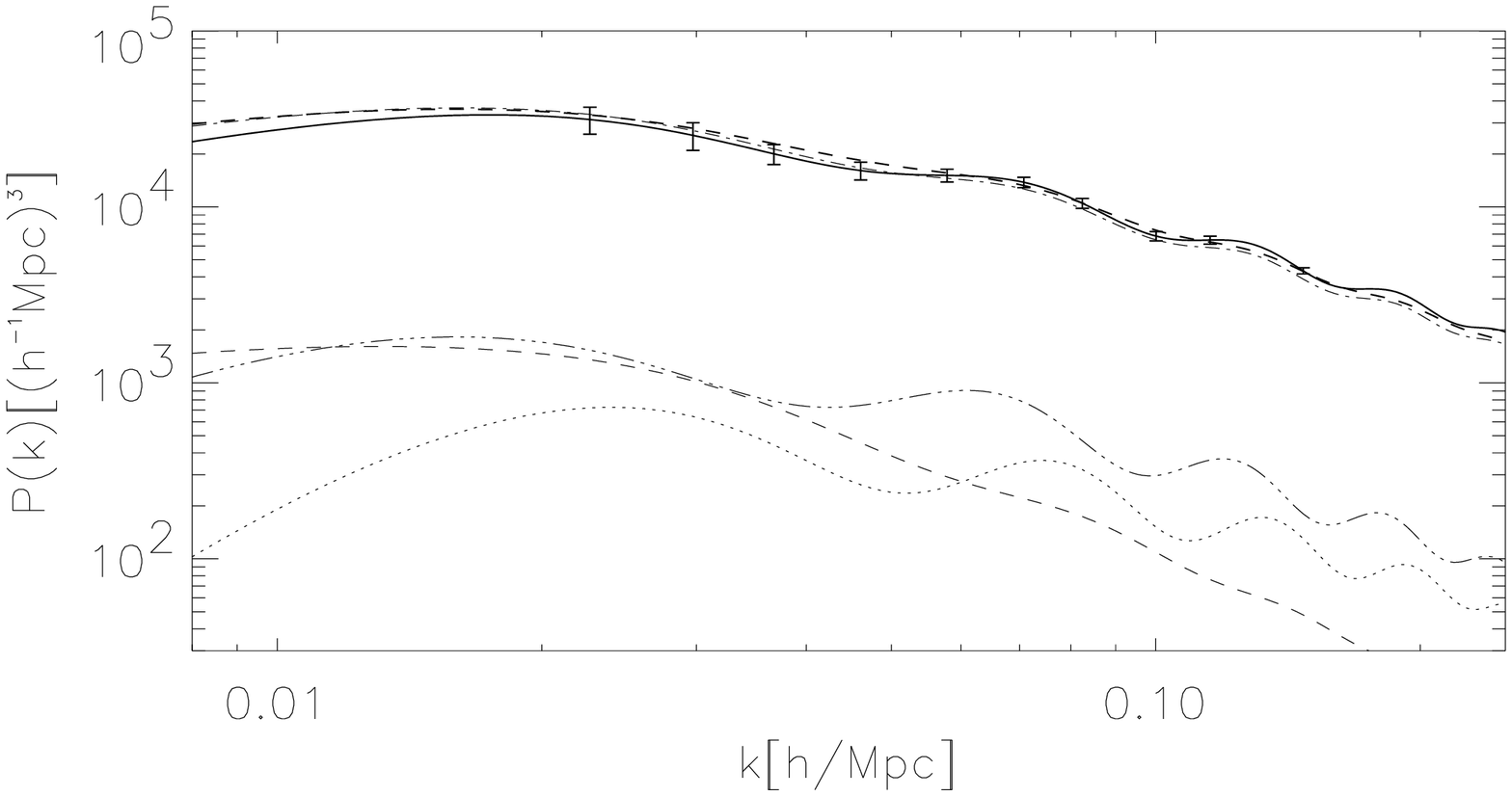,width=8.5cm,height=5.5cm}
\vskip +0.0in
\caption{
{\bf Power spectra of best-fit mixed models for three isocurvature modes.}
The angular power spectrum of CMB anisotropies, $C^{T}_\ell,$
the temperature-polarization cross correlation spectrum, $C^{TE}_\ell,$
and the galaxy power spectrum, $P(k),$ for the best-fit adiabatic model
(dashed line) and best-fit mixed model with $\fiso=60\%$
(solid line), with the auto-correlation contributions to the 
mixed model included. The CMB data
points are overplotted and the sizes of the 2dFGRS error bars are
indicated.
\label{cl_plot}}
\end{center}
\end{figure}

In Fig.~\ref{cl_plot} we plot the CMB and matter power spectra for an
adiabatic model and a mixed model with high likelihood, with
parameters ($\omega_b$, $\omega_c$, $\Omega_\Lambda$, $n_s$, $\tau$,
$\beta$, $\fiso$) =(0.041, 0.13, 0.75, 1.06, 0.28, 0.37, 0.60).
The adiabatic and mixed model spectra are essentially
indistinguishable, despite the adiabatic mode contributing only $40\%$ 
of the total rms power of the mixed model spectra. 

\begin{figure}[t]
\epsfig{file=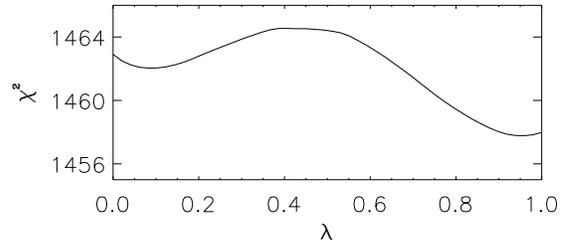,width=8cm,height=4cm}
\caption{
{\bf $\chi^2$ for models connecting adiabatic and mixed models}, 
calculated along a path 
connecting the best-fit adiabatic model ${\bf x}_1$ with the best-fit 
mixed model ${\bf x}_2$, calculated at
position  ${\bf x}= \lambda {\bf x}_2 + (1-\lambda) {\bf x}_1$. 
\label{like_path}}
\end{figure}

Although the posterior distributions for mixed models and pure
adiabatic models hardly overlap, the high-likelihood
models are connected in parameter space. That is to say, the
goodness-of-fit, $\chi^2,$ changes almost monotonically 
over a variation of parameters that interpolates linearly between
these models, as shown in Fig.~\ref{like_path}. This means that the
posterior distribution is not bimodal but rather flat between the high
likelihood adiabatic model and the high likelihood mixed model. A
principal component analysis of the covariance matrix of the
distribution, calculated about the high likelihood mixed model shown in 
Fig.~\ref{cl_plot},
indicates that the flattest direction ${\bf y}_0$ has an uncertainty
greater than $100\%$, with three eigen-directions having uncertainties 
greater than 50\%.

To elucidate what physical effects are responsible for these flat
directions, we search for degenerate directions that arise from smaller 
subsets of the full parameter set. We have identified two such
flat directions. The first is the degenerate direction described in
Section \ref{one_mode_section} involving seven parameters, which is
again present here. This flat direction accounts for the negative
correlations, $z_{\mdcs{AD}{CI}}$ and $z_{\mdcs{CI}{NID}},$ and the
positive correlation $z_{\mdcs{AD}{NID}}$. By performing a principal
component analysis of these seven parameters, we find a very similar
flat direction to the one shown in Fig.~\ref{degen_ci}.

\begin{figure}
\begin{center}
\epsfig{file=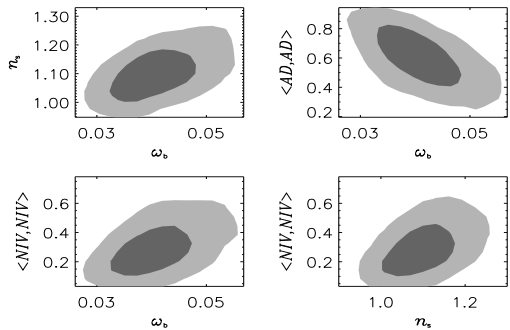,width=7cm,height=4.5cm}
\vskip +0.1in
\caption{
{\bf Two dimensional marginalized distributions.}
Correlations between the baryon density $\omega_b$, the spectral index 
$n_s$ and the contributions of the pure adiabatic and neutrino velocity 
isocurvature modes are illustrated. $\mdcs{I}{J}$ denotes $z_{\mdcs{I}{J}}$.
\label{twod_wb}}
\end{center}
\end{figure}

\begin{figure}[t]
\epsfig{file=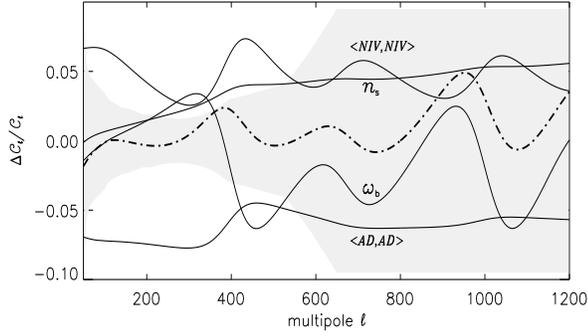,width=8cm,height=5cm}
\vskip +0.1in
\caption{
{\bf Degenerate direction in $\omega_b$, $n_s$, $A_{\mdcs{AD}{AD}}$, 
and $A_{\mdcs{NIV}{NIV}}$ }
The total derivative along the flattest degenerate direction
(dot-dashed) is decomposed into variations in the four parameters above (all
solid), which cancel to within the limits allowed by WMAP error bars
(shaded).
\label{degen_wb}} \vskip -0.2cm
\end{figure}

The second flat direction, that involves four parameters, is
associated with a large shift in the value of the baryon density. By
studying the two-dimensional marginalised distributions shown in
Fig.~\ref{twod_wb}, it is clear that the spectral index and the
adiabatic and neutrino isocurvature velocity mode contributions are strongly
correlated with the baryon density. Performing a principal component
analysis of the covariance matrix of this four-parameter set, reveals
a flat direction with an uncertainty of $50\%$. 
This cancellation is illustrated in
Fig.~\ref{degen_wb}, where the baryon density is increased by 8\%,
leading to the suppression (boosting) of even (odd)-numbered
adiabatic acoustic peaks. This shift is compensated by increasing the
power in the NIV mode by 29\%, reducing that in the AD mode by 13\%
and increasing the overall spectral index $n_s$ by 2\%, leaving the 
CMB temperature spectra unchanged to within
the WMAP experimental error bars. It is not
surprising that the neutrino velocity mode provides the oscillatory
cancellation as the baryon density increases, given that it is
approximately in phase with the adiabatic mode. 

We now consider the statistical significance of these results.
The mixed model reduces the total $\chi ^2$ by 5. Under the
approximation of a linear parametric dependence
(which in this case is probably not a very good approximation),
one would expect the reduction in $\chi ^2$ to be
governed by a $\chi ^2$ distribution with nine degrees of freedom
(i.e., equal to the number of extra parameters).
Consequently, a reduction by only 5 where simply fitting
the noise would lead to a reduction by 9, does not
constitute evidence in favor of adding the extra
parameters.

We conclude that although our results demonstrate that the data
allows for large amounts of isocurvature, we have not found any
evidence indicating that the models with more parameters offer
a statistically significantly better fit.
The same analysis may be applied to the additions of a single mode
and pairs of modes. Adding a single mode adds two parameters,
and we find a reduction of the total $\chi ^2$ by
1 for the NIV mode, and less than 1 for the CI and NID modes.
In none of these cases is the improvement in fit statistically
significant.
Similarly, for pairs of modes, five parameters are added, and
the total  $\chi ^2$ decreases by approximately 1, 1, and 2 for the
combinations CI+NID, CI+NIV, and NID+NIV, respectively.

It is possible in principle that a statistically significant
improvement in fit might be obtained if the exponents in the
power laws for the various modes were allowed to vary
independently and/or additional effects such as tensor
modes were allowed. We did not investigate this possibility.

\section{Sensitivity of posterior distributions to constraints and 
prior distributions.}
\label{Priors}

The results of the previous section were obtained using Bayes'
theorem
\ba
P_{posterior}(\alpha _{par}\vert {\bf x}_{data})\propto
L({\bf x}_{data}\vert {\bf \alpha }_{par})~
P_{prior}(\alpha _{par}),
\ea
where $\alpha _{par}$ spans the parameter space of models and 
$L({\bf x}_{data}\vert {\bf \alpha }_{par})$ is the likelihood
of the model labeled by ${\bf \alpha }_{par}$
given the CMB (or CMB+LSS) data indicated by ${\bf x}_{data}$,
subject to the prior distribution over the space of models
$P_{prior}(\alpha _{par}).$ The posterior distribution is 
computed using MCMC techniques outlined in \cite{jo} 
and references therein.  Certain prior distributions in the form of 
constraints, such as the positivity of various parameters that
physically cannot be negative or the positive definiteness of
the matrix-valued power spectrum, do not merit discussion. 
For most parameters uniform priors were assumed, following 
customary practice.

In this section we investigate the sensitivity of our 
conclusions (i.e., the posterior probability 
distribution) for those variables for which the choice of prior
was important. In particular, we investigate the sensitivity
to: (1) The allowed range for the reionization optical depth
$\tau ,$ (2) the priors for the redshift distortion 
parameter $\beta =b/\Omega ^{0.6}$ and the bias $b,$
(3) the prior for $\omega _b,$ that is whether a 
uniform prior is employed or whether information from
other independent determinations of  $\omega _b$ based
on nucleosynthesis are employed, and (4) the choice
of parameterization (or measure) for $A_{IJ}(k).$ 
In all cases we consider the $N=4$ example, including all possible 
isocurvature modes.

\subsection{Alternative Priors for Cosmological Parameters}

\subsubsection{Constraint on optical depth}

\begin{figure}
\epsfig{file=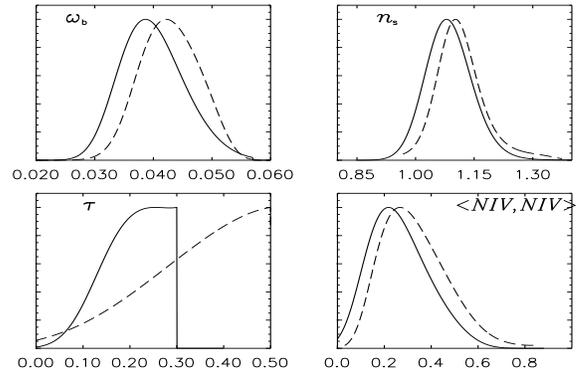,width=8cm,height=5cm}
\vskip 10pt
\caption{
{\bf Effect of changing the upper limit for $\tau $
from {\rm 0.3 (solid)} to {\rm 0.5 (dashed)}.} Only 
the posterior distributions for parameters where the 
difference is significant are shown; the distributions 
for the remaining parameters are 
unchanged from their $\tau < 0.3$ values.
\label{tau_prior}}
\end{figure}

Following the WMAP team analysis\cite{spergeletal} of 
pure adiabatic models, we limited the reionization 
optical depth to the range $0 \le \tau < 0.3$. When isocurvature modes 
were included, in particular the NIV mode, we observed that the distribution
for $\tau$ seemed to concentrate toward the endpoint at
$\tau =0.3.$ To investigate the sensitivity to the value
of this cutoff, we considered an alternative upper limit
at $\tau =0.5$. Fig.~\ref{tau_prior} compares the posterior 
distributions for these two upper
bounds, showing those parameters for which
the difference is greatest. Even with the upper bound raised to
$\tau =0.5$ the posterior distribution still accumulates there. 
However, the effect on other parameters is mild, with only a 
slightly larger isocurvature contribution at $\fiso=0.60\pm0.10$ 
permitted.

\subsubsection{Priors from galaxy redshift survey data}

We have included data on the galaxy power spectrum from the 2dFGRS
survey, incorporating a Gaussian prior for the redshift
distortion parameter with $\beta =0.43 \pm 0.077,$ that is
derived from a measurement of the redshift distortion effect observed
in the galaxy survey \cite{twodf,verde}. The WMAP
team \cite{spergeletal} have argued for an independent prior on the
bias of $b=1.04\pm 0.11,$ obtained by computing the bispectrum of the
2dFGRS survey \cite{verde_bias}. In the previous section we did 
not impose this prior on
the bias since it was obtained under the assumption of 
Gaussian initial fluctuations. Furthermore, an implicit prior is placed on 
$\Omega_m$ (and therefore $\Omega_\Lambda$) when it is combined with the 
prior on $\beta$. 
We found a higher value for the bias, $b = 1.3 \pm 0.2$, obtained 
from constraining three isocurvature modes with
the CMB+LSS dataset. A larger bias generally allows for a
larger isocurvature fraction, so we now include the additional 
determination of $b$ to determine its effect
on the allowed isocurvature contributions. The
effect is small, with parameter value distributions virtually the same
and the isocurvature contribution decreasing only slightly from
$\fiso=0.57$ to $0.53$.

\begin{figure}
\epsfig{file=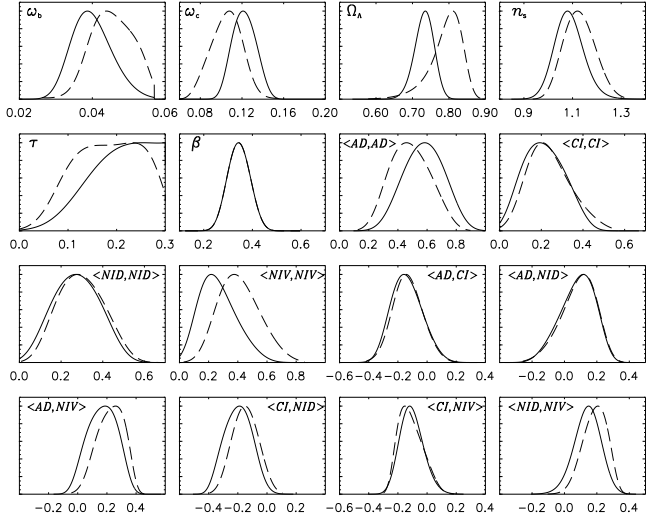,width=8.5cm,height=7cm}
\vskip 0.1in
\caption{
{\bf Effect of discarding LSS amplitude information.}
Marginalized parameter distributions obtained using only the
shape information from the LSS data (dashed) compared 
to those distributions obtained using the prior
$\beta=0.43\pm0.077$ (solid).
\label{freebias_prior}}
\end{figure}

We have also investigated the role of the prior on $\beta$ in
constraining the ratio of the galaxy power spectrum to the matter
power spectrum, $f(b,\beta)$. By relaxing the prior on $\beta$, we
allow $f$ to vary freely so that only information on the shape of the
power spectrum is utilised. The resulting parameter distributions are
shown in Fig.~\ref{freebias_prior}, where $f$ has been optimised at
each step in the chain. We observe that larger isocurvature fractions,
$\fiso=0.65\pm0.09$ are allowed.
In particular the NIV mode contribution increases to 
$z_{\mdcs{NIV}{NIV}}=0.40\pm0.15$. 
We have shown that such an increase is correlated with a
reduction in the amplitude of the adiabatic mode. Such models were 
previously disfavoured by the prior on $\beta$ because of their 
resulting low matter power spectrum amplitudes, but are 
permitted when the
normalisation of the matter power spectrum is free to vary. More 
models along the flat NIV direction are explored, thereby
permitting larger values of $n_s$ and $\omega_b$. We also observe
weaker constraints on $\omega_c$ and $\Omega_\Lambda$. This test
demonstrates that the redshift distortion measurement is very
informative.

\subsubsection{Prior on baryon density}

\begin{figure}[t!]
\vskip 0.1in
\epsfig{file=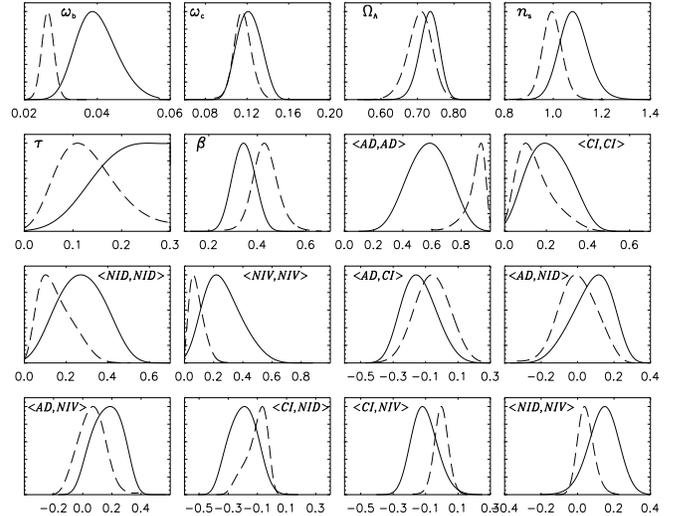,width=8.5cm,height=7cm}
\vskip 0.1in
\caption{
{\bf Effect of incorporating the nucleosynthesis
determination of $\omega _b.$}
The curves show the posterior distributions from the
CMB+LSS data with a flat
prior for $\omega_b$ (shown in solid) and 
a Gaussian prior reflecting the determination
$\omega_b=0.022\pm0.002$ (shown as dashed).
\label{bbn_prior}}
\end{figure}

In the previous section, when three possibly correlated isocurvature
modes were allowed, we found $\omega_b = 0.041\pm 0.006$ 
for the baryon density, a value discrepant with nucleosythesis
based determinations of $\omega_b,$ which indicate 
$\omega_b = 0.022 \pm 0.002$ \cite{bbn}.
We investigate the effect of replacing the uniform prior
on $\omega_b$ with a prior that incorporates this 
complementary information.
Fig.~\ref{bbn_prior} shows the posterior distributions 
with and without the nucleosythesis determination.
The median values and 68\% confidence levels are shown in Table
\ref{table_ad_3iso_priors}. The posterior for the baryon density 
overlaps with the prior nucleosythesis distribution but values as large as
$\omega_b=0.030$ are still permitted. There is a large reduction in
the isocurvature contribution, in particular the NIV mode and its
correlations, with the overall non-adiabatic fraction reduced to a
median value of $\fiso=0.31$. The cosmological parameters 
correlated with the NIV amplitude, such as $n_s$, $\tau$
and $\beta$ 
shift back toward their pure adiabatic values.

\begin{table}[t!]
\begin{minipage}[h]{.48\textwidth}
\begin{center}
\begin{tabular}[h]{lcc}
\hline\hline
& \hspace{-4mm} \mbox{\fns CMB+LSS} 
& \mbox{\fns CMB+LSS} \\ 
& \hspace{-4mm} \mbox{\fns (BBN)} 
& \mbox{\fns (standard)} \\ \hline
\hspace{0.25cm}$\omega_b$ & \hspace{-4mm}
\mbs{$0.026\pm 0.002$}   &
\mbs{$0.041\pm 0.006$} \\
\hspace{0.25cm}$\omega_c$ & \hspace{-4mm}
\mbs{$0.116\pm 0.008$}  & \mbs{$0.12\pm0.01$} \\
\hspace{0.25cm}$\Omega_\Lambda$ & \hspace{-4mm}
\mbs{$0.71\pm 0.03$} 
& \mbs{$0.74\pm0.03$} \\
 \hspace{0.25cm}$n_s$ & \hspace{-4mm}
\mbs{$0.99\pm0.04$}  & \mbs{$1.10\pm0.06$} \\
 \hspace{0.25cm}$\tau$ & \hspace{-4mm} \mbs{$0.12^{+0.07}_{-0.05}$}
 &  \mbs{$0.22\pm0.07$} \\
 \hspace{0.25cm}$\beta$ & \hspace{-5mm}
\mbs{$0.43\pm0.05$}  & \mbs{$0.35\pm0.05$} \\
\\
$z_{\mdcs{AD}{AD}}$ & 
\mbs{$0.91\,$}$^{+\,0.05}_{-\,0.08}$ 
& \mbs{$0.61\pm{0.15}$} \\
$z_{\mdcs{CI}{CI}}$ &
\mbs{$0.13\,$}$^{+\,0.10}_{-\,0.07}$
& \mbs{$0.23\pm0.11$} \\
$z_{\mdcs{NID}{NID}}$ & 
\mbs{$0.14^{+0.10}_{-0.06}$}
& \mbs{$0.30\pm0.12$} \\
$z_{\mdcs{NIV}{NIV}}$ &
\mbs{$0.08\,$}$^{+\,0.06}_{-\,0.04}$
& \mbs{$0.28\,$}$^{+\,0.14}_{-\,0.11}$ \\
$z_{\mdcs{AD}{CI}}$ &
\mbs{$-0.06\pm0.11$} 
& \mbs{$-0.12\,$}$^{+\,0.12}_{-\,0.10}$ \\
$z_{\mdcs{AD}{NID}}$  &
\mbs{$0.0008\pm0.11$} 
&\mbs{$0.11\,$}$^{+\,0.10}_{-\,0.12}$ \\
$z_{\mdcs{AD}{NIV}}$ &
\mbs{$0.06\pm0.09$} 
&\mbs{$0.19\pm0.11$} \\
$z_{\mdcs{CI}{NID}}$  &
\mbs{$-0.10\pm0.09$} 
&\mbs{$-0.18\pm0.10$} \\
$z_{\mdcs{CI}{NIV}}$ &
\mbs{$-0.003\pm0.04$}  &\mbs{$-0.09\pm0.08$} \\
$z_{\mdcs{NID}{NIV}}$ &
\mbs{$0.04\pm0.05$} 
&\mbs{$0.16\pm0.08$} \\
\\
  $z_{\mbox{\fnsc ISO}}$  &\mbs{$0.41\pm0.14$} &
\mbs{$0.79\pm^{0.09}_{ 0.13}$}\\
$f_{\mbox{\fnsc ISO}}$ & \mbs{$0.31\pm0.09$}   &
\mbs{$0.57\pm 0.09$} \\
\\
$\Omega_m$ &  \mbs{$0.29\pm0.03$}  &  \mbs{$0.26\pm0.03$} \\
$h$ & \mbs{$0.70\pm 0.04$}  
& \mbs{$0.80\pm 0.05$}\\
$b$ & \mbs{$1.1\pm0.1$}& \mbs{$1.3\pm 0.2$}  \\
\hline
\end{tabular}
\end{center}
\end{minipage}\hfill

\vskip 0.1in
\caption{ {\bf Effect of incorporating the nucleosynthesis
determination of $\omega _b.$}
Median parameter values and $68\%$ confidence intervals as in 
Table~\ref{table_ad_1iso} for mixed models with a
nucleosynthesis prior (BBN, 1st column), compared to the
standard results with a broad uniform prior (2nd column). 
The CMB+LSS dataset is used throughout.
\label{table_ad_3iso_priors}}
\vskip -0.2in
\end{table}

\subsection{Dependence on mode parameterization}

We examine the sensitivity of
the posterior distribution to the choice of
prior distribution for the coefficients of the 
matrix ${\cal A}_{IJ}=A_{IJ}~k^{n_S}.$ As detailed in 
section IIB, after normalizing 
the modes according to their contribution to 
the total CMB power from $\ell =2$ through
$\ell =2000,$ we parameterized 
$A_{IJ}$ as $A_{IJ} \propto z_{IJ}$
where the constraint $\| z \| =1$ was imposed
and the uniform spherical measure on the 
resulting sphere was assumed. A uniform prior
for the constant of proportionality was assumed.

\begin{figure}[t!]
\epsfig{file=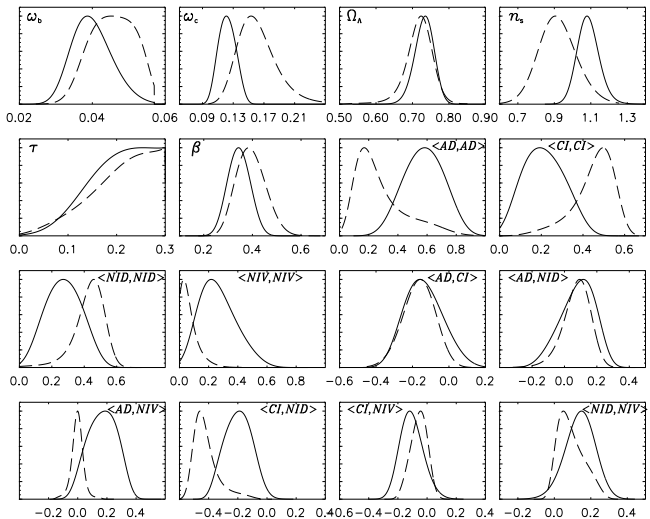,width=8.5cm,height=7cm}
\vskip 0.1in
\caption{{\bf Alternative prior uniform in the coefficients}
$A_{IJ}.$ The posterior distributions with this new prior
(dashed) are compared to those of the 
`standard' prior described in Section II (solid).
\label{abs_dist}}
\end{figure}

We now consider directly sampling a uniform distribution 
in the coefficients
$A_{IJ}.$ This alternative prior makes a large 
difference for the case of three isocurvature 
modes, for which the allowed isocurvature 
rises from $\fiso=0.57$ with the old
prior to $\fiso=0.81$ with the new prior. 
The posterior distributions are modified
significantly, as indicated in 
Fig.~\ref{abs_dist}.

We can understand this effect by observing that the constant 
of proportionality in eqn. \ref{eqn:amplitude}, which we now label 
$A^\prime$, can be written as 
$(\sum_{IJ} A_{IJ}^2)^{1/2}.$
Under this new parameterization where we sample uniformly 
in $A_{IJ}$, we therefore replace
the uniform prior on ${A^\prime}$ with a prior so that
$P(A^\prime) \propto (A^{\prime})^ \alpha $ where $\alpha =N(N+1)/2-1.$
This favours models with high values for
${A^\prime }$, corresponding to those with large positive 
and negative 
isocurvature contributions $A_{IJ},$ which may cancel out to fit the data.
As discussed in the previous
section, such models may be formed by moving along a degenerate 
direction consisting of 
the auto- and cross-correlations formed by the CI and 
NID modes with the
adiabatic mode, and the spectral index $n_s.$ 
A similar direction was illustrated in 
Fig.~\ref{degen_ci}.
These models account for the increased contribution of 
both the CI and NID mode
auto- and cross-correlations, and hence 
the reduction of the relative
AD and NIV mode contributions.
While the likelihood of such models with very high 
isocurvature is relatively low, the phase-space 
effect due to the modified prior 
boosts their posterior probability.

\begin{figure}[t!]
\epsfig{file=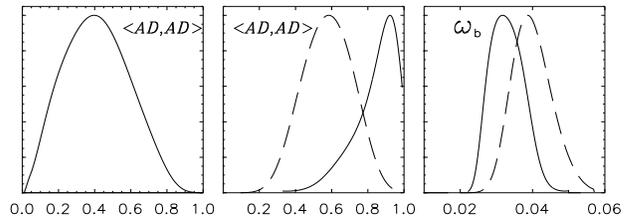,width=8cm,height=3cm}\vskip +0.2in
\caption{
{\bf Effect of reweighting to correct for bias against 
pure models.} The left panel shows what  
posterior for the adiabatic mode
contribution $z_{\mdcs{AD}{AD}}$ would be in the absence of data.
In the center panel we show how the posterior (from the prior in
section IIB) (dashed) shifts (to the solid) curve)
from a re-weighting that would give
a flat posterior in the absence of data. The right panel shows
the corresponding shift in $\omega _b$ resulting from the same
reweighting.
\label{unif_zad}}
\end{figure}

Both of the previous priors disfavor models where any 
one mode dominates over the others. This is because of the 
product factor in the distribution given 
in eqn.~(\ref{eqn:eval}).
For the prior presented in section IIB, taking the limit
$z_{\mdcs{AD}{AD}}\to 1$ causes one of the eigenvalues of $z_{IJ}$ to approach
one while the other $(N-1)$ eigenvalues approach zero.
More quantitatively, if $z_{\mdcs{AD}{AD}}=1-\epsilon ^2, $ 
the $(N-1)$ small eigenvalues cluster within $\epsilon $
of zero, giving a factor proportional to $\epsilon ^{N(N-1)/2}$
in the product in eqn.~(\ref{eqn:eval}). This gives a prior
density for $z_{\mdcs{AD}{AD}},$ marginalized
with respect to the other parameters, 
proportional to $\epsilon ^{N(N-1)/2-1}$
in the neighborhood of $z_{\mdcs{AD}{AD}}.$ Consequently,
even if the data indicated that all models where equally
likely---that is, if the data was non-informative---our
posterior would have a zero of the same order
at the endpoint $z_{\mdcs{AD}{AD}}=1$! 

We may correct for such bias, or estimate its effect, by 
rescaling the prior so that
in the absence of data the posterior in  $z_{\mdcs{AD}{AD}}$
would be flat, as shown in Fig.~\ref{unif_zad}.
The left panel shows the posterior for $z_{\mdcs{AD}{AD}}$
in the absence of data for $N=4$ (i.e., adiabatic +
three isocurvature modes). In the middle panel
the (dashed) curve shows the posterior for the prior
of section IIB. When this is rescaled (by dividing by 
the posterior with no data and then renormalizing)
the posterior represented by the solid curve
results. The right panel shows the results of
this reweighting on $\omega _b.$
With this re-weighting
the adiabatic fraction increases to
$z_{\mdcs{AD}{AD}}=0.87^{+0.08}_{-0.14}$, with a lower isocurvature
fraction, $\fiso=0.36\pm0.12.$ 

\section{Discussion}
\label{Discussion}

We have presented a framework within which to consider
correlated mixtures of adiabatic and isocurvature perturbations and
to obtain simultaneously
more model independent constraints on cosmological
parameters as determined
from CMB and large-scale structure datasets. 
Applying these methods to recent CMB data,
from WMAP and several small-scale experiments, indicates that current
constraints on a single correlated isocurvature mode are strong ($\approx 
10\%$). These limits, however, degrade substantially when two or three
correlated isocurvature modes are allowed, with non-adiabatic
fractions as large as $60\%$ possible. Including 
large-scale structure data from the 2dF survey only slightly modifies
these constraints. 

The parameters significantly modified by including
correlated isocurvature initial conditions are $\omega_b, \tau, \beta$
and $n_s,$ with the values of the baryon density values twice as large as
in the pure adiabatic case not ruled out. 
These parameters are strongly correlated
with the isocurvature amplitudes and for the current 
data  are in fact degenerate along a well-defined direction in
parameter space. 
Consequently, the likelihood function is not strongly
peaked around a given model but relatively flat 
in a region interpolating between a set of mixed
models and the pure adiabatic model. 
For many of the allowed models
the large isocurvature fraction did not necessarily 
accompany a substantial decrease in the adiabatic power. 
This is possible because of
interference phenomena as we explained. 
While we find that large isocurvature fractions are allowed 
we do not find evidence that the inclusion of such modes
provides a statistically significant better fit to the present 
data. 
\\
{\it Acknowledgments}: K.M. and J.D. acknowledge the support of 
PPARC. M.B. thanks Mr D.~Avery for financial support. 
P.G.F. thanks the Royal Society. C.S. is supported by a Leverhulme 
trust grant. 
We thank  R.~Durrer, U.~Seljak, D.~Spergel and R.~Trotta for 
useful discussions.

\appendix
\section{Rapid computation of cosmological models using DASh}
\label{Software}

\subsection{Motivation and Description}

Cosmological parameter estimation currently requires
calculating at least $10^6$ models when a 
typical 6-dimensional parameter space is
sampled uniformly. 
Using an MCMC sampler, which favors the region of
high likelihood, reduces 
the number of computed models to $\approx 10^4$ 
for the
same parameter space. Future parameter
estimation studies will aim to measure many more parameters as the
quality of CMB temperature and polarisation data improves, thereby
increasing the computational cost. By including
non-adiabatic initial conditions in this study,
we have enlarged the parameter space
to 16 dimensions, requiring computing many more models.
The development of software that speeds up model computation 
is crucial for sampling such large parameter spaces.

Current codes take between 30 and 60 seconds to compute a single
model, which is prohibitively long for sampling a
large dimensional parameter space. One requires 
a method that both improves the precision of
the computed $C_\ell$ spectrum as data becomes more accurate 
without sacrificing speed. The method adopted here and extended 
is the Davis Anisotropy Shortcut (DASh) \cite{DASh}. 
With DASh, the accuracy depends primarily on the fineness
of the grid, which depends on the size of available
fast memory. Consequently, accuracy may be increased without
sacrificing speed. 

DASh was initially developed as a fast method for predicting the
angular power spectrum given a set of cosmological parameters
(the physical baryon density $\omega _b$, physical matter density 
$\omega _m$, relative cosmological constant density $\Omega_\Lambda$, 
relative
curvature density $\Omega_K$ and reionization optical depth $\tau$)
and an arbitrary primordial power spectrum. The method relied on the
construction of grids: a {low-$\ell$ grid}
$${\cal G}_\ell^L[w_b,w_m,\Omega_\Lambda,
\Omega_K], \qquad 2 \leq \ell \leq \ell_t$$ 
for an upper
threshold $\ell_{t}[\Omega_\Lambda,\Omega_K]$; a {high-$\ell$
grid}, 
$${\cal G}_\ell^H[w_b,w_m], \qquad \ell>\ell_{t};$$ 
a {polarization grid} 
$${\cal G}_\ell^P[w_b,w_m];$$  
and a reionization grid 
$${\cal G}_\ell^R[w_b,w_m,\tau], \qquad 2 \leq \ell \leq \ell_R$$ 
for an upper threshold $\ell_{R}$. 
At each vertex of the low-$\ell$ and high-$\ell$ grids 
the temperature transfer function $\Delta_\ell^T(k)$ 
was pre-computed and stored. Similarly, for the
polarization grid the polarization transfer functions,
$\Delta_\ell^E(k),$ were pre-computed and stored. For the reionization
grid, a function of $C_\ell^{T}[\tau]$ and $C_\ell^{T}[\tau=0]$ was
stored at each grid point. After the initial construction of the grids, an
arbitrary $C_\ell$ spectrum could be calculated in roughly 1 to 2 seconds, 
as long as its parameters were within the
parameter range of the grid. 

The $C_\ell^{T}$, $C_\ell^{TE},$ and $C_\ell^{E}$
spectra were calculated by
interpolating the transfer functions,
$\Delta_\ell^T(k)$ and $\Delta_\ell^E(k)$ from the values at the
nearest vertices on the relevant grids.
A linear
interpolation of the quantities, $|\Delta_\ell^T(k)|^2$,
$\Delta_\ell^T(k)\Delta_\ell^E(k)$ and $|\Delta_\ell^E(k)|^2$ was
performed before integrating over $k$ to obtain the low $C_\ell^{T}$
spectrum, and the high $C_\ell^{T}$, $C_\ell^{TE}$ and $C_\ell^{E}$
spectra for $\tau=0$. The high $C_\ell^{T}$, $C_\ell^{TE}$ and
$C_\ell^{E}$ spectra were then rescaled in $\ell $ to take into
account  the modified angular diameter distance to recombination. 
The low and high $\ell $ temperature 
spectra were then combined to obtain the 
$C_\ell^{T}$, $C_\ell^{TE}$ and $C_\ell^{E}$ spectra for
$\tau=0$. The final spectrum was calculated by multiplying by a
reionization correction envelope obtained from the reionization 
grid for $\ell\leq \ell_R$ and a suppression factor $e^{-2\tau}$ 
for $\ell> \ell_R$ in the temperature case. 
For the temperature-polarization
cross-correlation and polarization auto-correlation, 
a fitting
function for the reionization bumps was used. More details of the
initial release \cite{DASh_web} of DASh can be found
in \cite{DASh}. Current alternative methods to perform rapid CMB model
computation or to sample parameter space rapidly include \cite{KMJ}
and \cite{STWZ}.

\begin{figure*}[t!]
\centerline{
\hbox{\epsfig{file=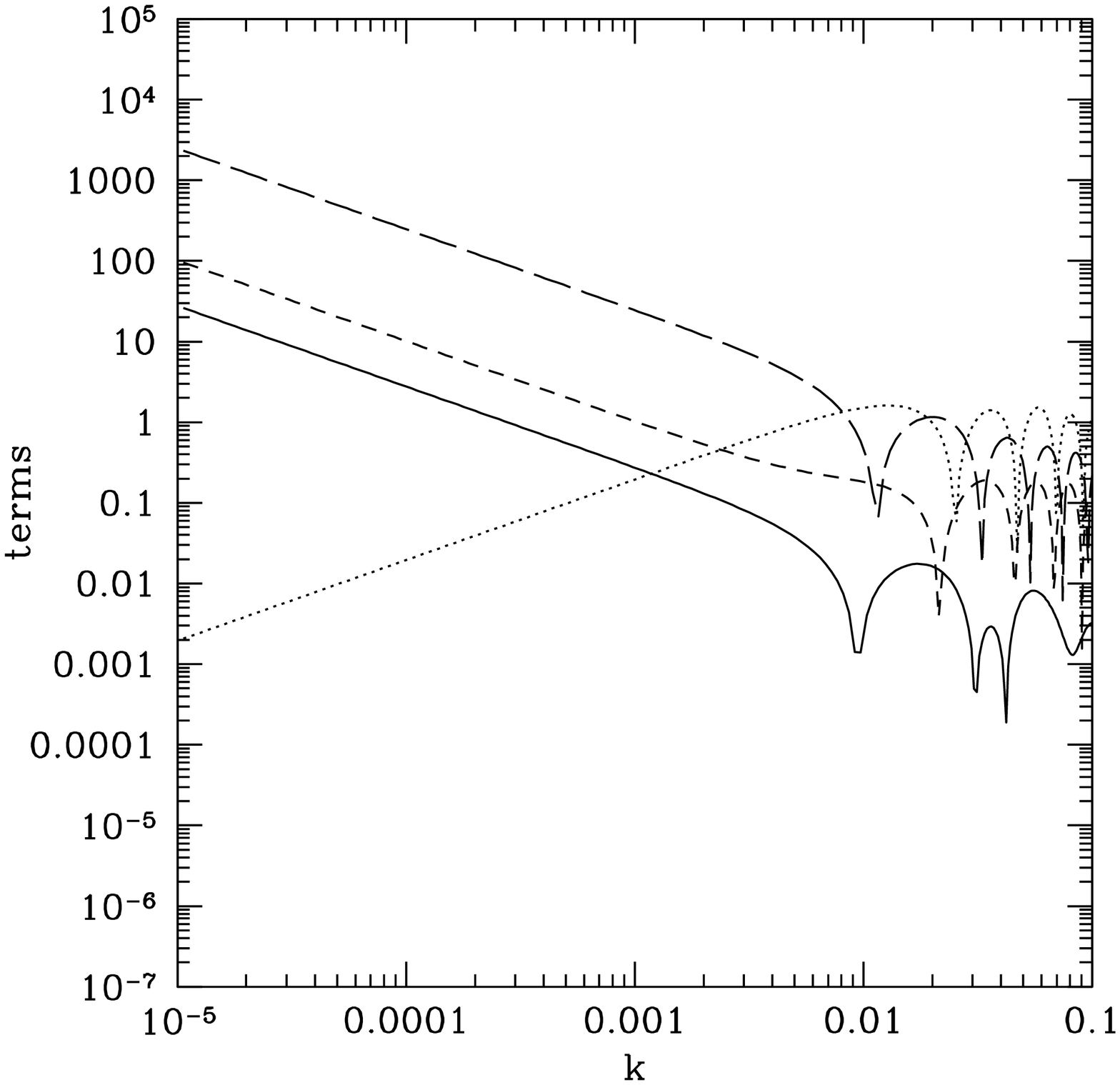,width=8.5cm,height=6.5cm}\hfill
\epsfig{file=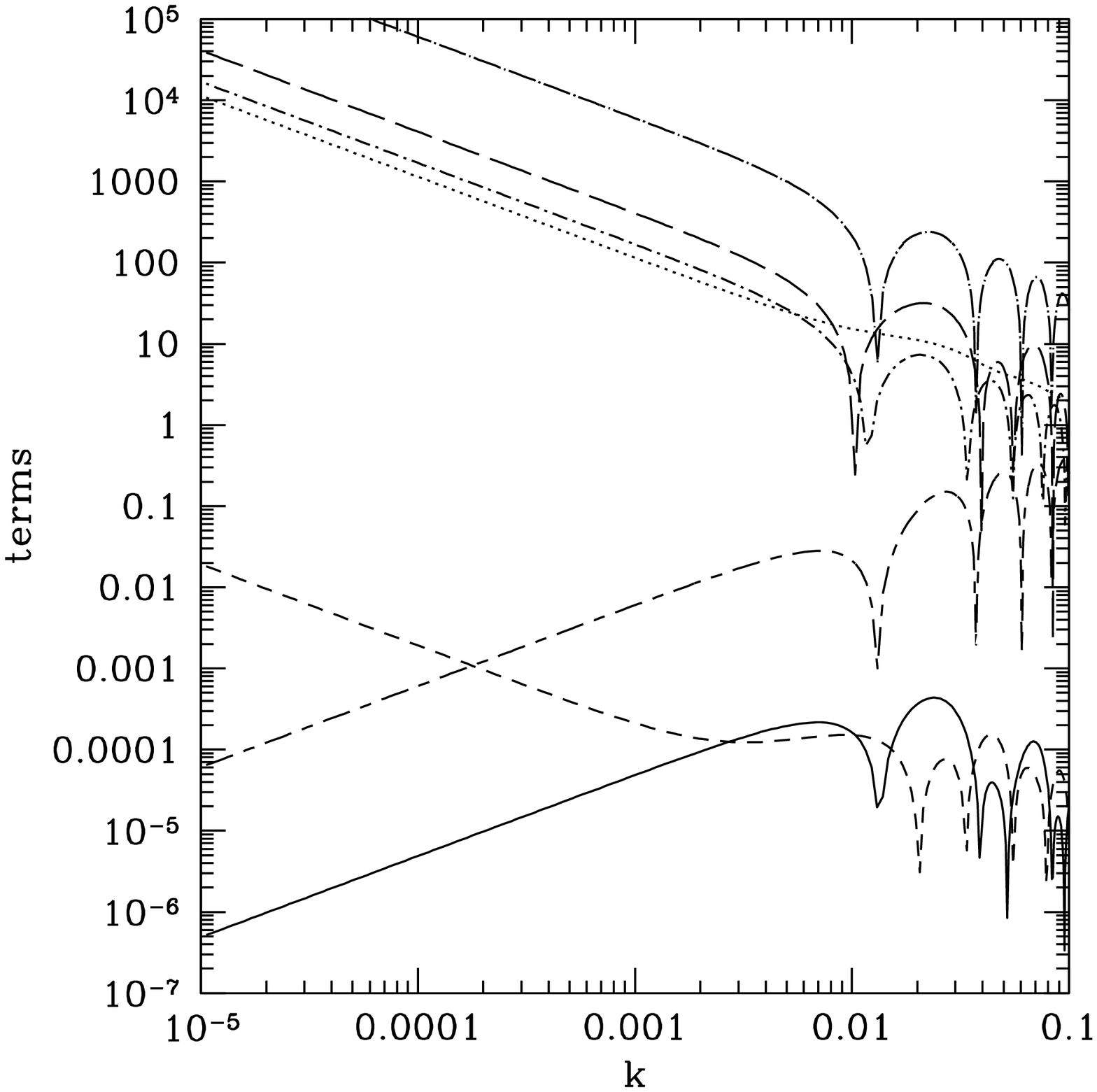,width=8.5cm,height=6.5cm}\hfill}}
\vskip +0.15in
\caption{The different terms in the Line-of-Sight
integral (eqn.~\ref{eq:Dlk_T}) separated into ones that are positive [left]
and negative [right] in the limit $k\rightarrow 0$. The positive terms
are : $\dot{\alpha}$ (solid), $\delta_\gamma$ (dotted),
$\frac{\dot{\theta}_b}{k^2}$ (dashed) and $\frac{\ddot{\Pi}}{k^2}$ (long
dashed). The negative terms are : $\dot{\eta}$ (solid),
$\alpha$ (dotted), $\ddot{\alpha}$ (dashed), $\frac{\theta_b}{k^2}$ (long
dashed), $\frac{\dot{\Pi}}{k^2}$ (dot-dashed),
$\frac{\Pi}{k^2}$ (dot-long dashed) and $\Pi$ (short-long
dashed). Clearly $\delta_\gamma$, $\dot{\eta}$ and $\Pi$ are finite
($\propto k$) as $k\rightarrow 0$ whereas the rest diverge as
$\frac{1}{k}$. }
\label{fig:div}
\end{figure*}

\subsection{Code extensions}
We extended the initial DASh version by including 
all isocurvature modes as in \cite{bmt}. In the new version, the
low-$\ell$, high-$\ell$ and polarization grids were created for each
pure mode. The cross-correlations between modes were calculated at
the interpolation stage. We have replaced the fitting functions
for the reionization bumps with two more reionization grids in $C_\ell^{TE}$ and
$C_\ell^{E},$ for greater accuracy and because the isocurvature mode
bumps had very different shapes to the adiabatic mode bumps. The three
reionization grids were therefore calculated for the auto-correlation and
cross-correlation modes.

We have also added the pre-computation of a matter transfer function
$T(k)$ grid 
$${\cal G}^T(k)[\omega _b,\omega _m]$$ 
for each pure mode with $\Omega_\Lambda
= 0.6$ for the range in $k$ space covered by the 2dFGRS
dataset \cite{2dF_percival}. The matter power spectrum was then
computed by interpolating $T(k)$ for each mode, multiplying by the
ratio of growth functions for $\Omega_\Lambda \ne 0.6$ and finally
obtaining $P(k)$ by summing over modes. The computation of a single
$P(k)$ is virtually instantaneous (after pre-computation of the grid).

Finally, due to the ${1}/{k}$ divergences that appear in some
terms (e.g., the Newtonian potentials) of the line-of-sight (LOS)
integral \cite{LOS} for the NIV mode, special care must be taken
when pre-computing the grids, as this introduces a numerical error as
large as $5\%$ to $10\%$ for $\ell<10$ which could be amplified to 
as much as $50\%$ after the interpolation. We discuss these details next.

\subsection{Neutrino isocurvature velocity mode}

For the neutrino isocurvature velocity mode, the Newtonian
potentials which are defined in the conformal Newtonian gauge (see for
example \cite{MB}), diverge as ${1}/{k}$ and
${1}/{t}$ where $t$ denotes 
conformal time. To first order they are 
\begin{eqnarray}
   \Psi &=&  -\frac{4R_\nu}{15 + 4R_\nu} \; \frac{1}{kt} + \ldots \\
   \Phi &=& +\frac{4R_\nu}{15 + 4R_\nu} \; \frac{1}{kt} + \ldots \\
\end{eqnarray}
where $R_\nu$ is a constant defined in \cite{bmt}. These divergences
have created confusion in the past and have led some authors to
discard this mode as unphysical. The singularities above, however, are
coordinate singularities, and disappear in
synchronous gauge.
All physical observables such as 
$C_\ell$s are finite for the NIV mode.

\begin{figure*}[t!]
\centerline{\hbox{\epsfig{file=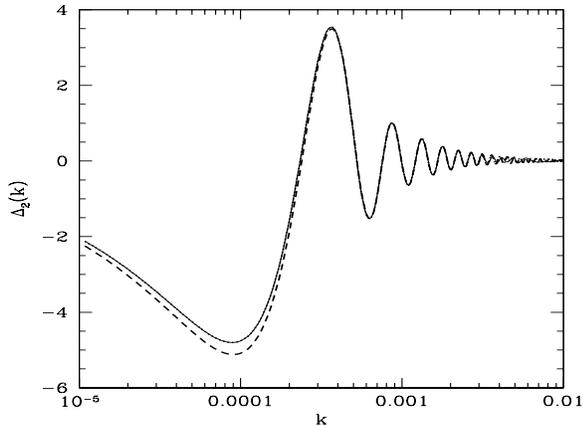,width=8cm,height=6cm}\hspace{1cm}
\epsfig{file=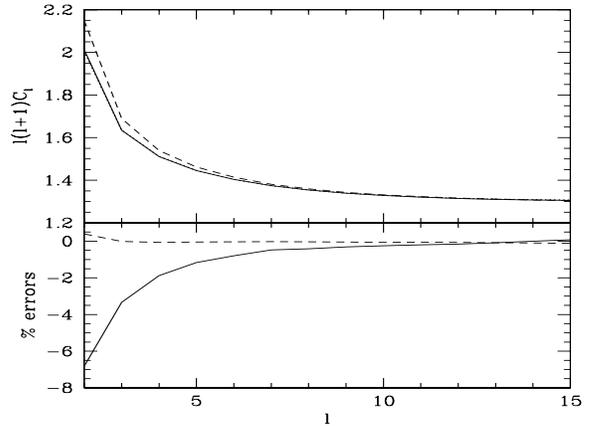,width=8cm,height=6cm}}}
\vskip +0.15in
\caption{
{\bf Accuracy of the computation of the temperature transfer function.}
[Left] The temperature transfer function $\Delta_2(k)$ calculated
using a
Boltzmann code (solid), with the line-of-sight integral (dashed) and with the
integration by parts (dotted). [Upper Right] The angular power spectrum
$C_\ell$ for a Boltzmann code (solid), LOS (dashed) and by
parts (dotted). [Lower Right] Percent difference between LOS and
Boltzmann (solid) and by parts and Boltzmann (dashed).}
\label{fig:cl_NIV}
\end{figure*}

Let $t_0$ be the conformal time today, $t_i$ an initial conformal
time deep into the radiation era, $x\equiv k(t_0 - t),$ and 
$g(t)=\dot{\tau}(t)e^{\tau(t)}$ the visibility function. The
optical depth back to $t$ is defined as $\tau (t)= \int_{t_0}^t
dt \;\; \dot{\tau}$. 
We use the conventions
and perturbation variables of \cite{MB}; $h$ and
$\eta$ are the synchronous metric perturbations, $\delta_\gamma$ 
the photon density contrast, $\theta_b$ the baryon velocity
divergence, $\Pi = F_2 + G_0 + G_2$ the polarization source term
with $F_\ell$ and $G_\ell$ the sum and difference of the two photon
polarizations, and $\alpha \equiv \frac{1}{2k^2}(\dot{h} +
6\dot{\eta})$ the gauge transformation variable. 
For the neutrino isocurvature velocity mode, at early times 
and on large scales the above variables are given to leading order in $t$ by
\begin{eqnarray}
    \delta_\gamma \propto  kt + \ldots ,\quad
    \theta_b \propto   k  + \ldots  ,\quad
    \eta  \propto  kt + \ldots  ,\quad \nonumber\\
    \alpha  \propto  \frac{1}{k} +  \ldots ,\quad
    \Pi  \propto   kt + \ldots .
\end{eqnarray}

The line-of-sight integral \cite{LOS} for the temperature transfer
function is given
by 
\begin{eqnarray} 
\Delta_\ell(k) =&&\int_{t_i}^{t_0}  \;  dt \; j_\ell(x) \; \Bigg\{ \, 
e^{\tau}(\dot{\eta} + \ddot{\alpha}) \nonumber\\
&& + \,  g(t)\left[ \frac{1}{4}\delta_\gamma + \frac{\dot{\theta}_b}{k^2} + 
2 \dot{\alpha} + \frac{1}{16}\Pi 
 \frac{3}{16k^2}\ddot{\Pi} \right]  \nonumber\\
&& + \, \dot{g}(t)\left[ \frac{\theta_b}{k^2} +
\alpha + \frac{3}{8k^2}\dot{\Pi} \right] +  \ddot{g}(t) \frac{3}{16k^2}\Pi
\Bigg\}
  \label{eq:Dlk_T}
\end{eqnarray}

Most terms in the above equation diverge as ${1}/{k}$ as shown
in Fig.~\ref{fig:div}. The transfer functions, $\Delta_\ell(k),$
however, are finite, since the Bessel functions are well approximated by
$j_\ell(kt) \propto (kt)^\ell$ at small $k,$ which is sufficient to
cancel the ${1}/{k}$ divergence. This means that if
$\Delta_\ell(k)$ is calculated using an exact Boltzmann code, it will
be finite. Using the LOS integral above, however, one should worry
that in computing $\Delta_\ell(k)$ the divergent terms will cause
numerical noise on large scales, because of the subtraction of large
numbers of (nearly) equal magnitude. This indeed occurs, with resulting
errors ranging from $5\%$ to $10\%$. Such errors are less 
a problem for codes based
entirely on the LOS method because the errors lie below the cosmic
variance. Using the DASh code, however, the interpolation between models
with this type of random error boosts the final error to
as much as $30\%$ to $50\%$ at $\ell<10$ in some cases.

In order to eliminate the above numerical error, one can integrate 
eqn.~(\ref{eq:Dlk_T}) by parts, so that
\begin{eqnarray}
\Delta_\ell(k) = \int_{t_i}^{t_0} \;  dt \; \Bigg\{ && j_\ell(x) \; 
\left[ e^{\tau}\dot{\eta} +
  \frac{1}{4}g(t)\left(\delta_\gamma + \frac{1}{4}\Pi \right) \right] \nonumber \\
&&
+\, \frac{dj_\ell}{dx} \left[
 e^{\tau}k\dot{\alpha} + \, \dot{g}(t) \frac{3}{16k}\Pi \,  \right.\nonumber \\
  && \qquad +\,  
\left.  g(t)\left(k\alpha +
    \frac{\theta_b}{k} + \frac{3}{16k}\dot{\Pi} \right)
 \right] \Bigg\}\nonumber
  \label{eq:Dlk_T_v2}
\end{eqnarray}
A quick check demonstrates that all the terms in the above equation
are finite as $k\rightarrow 0$, which gives stable and accurate
results when integrated numerically. We compare the different
integration methods in Fig.~\ref{fig:cl_NIV}.

\begin{figure}
\center
\epsfig{file=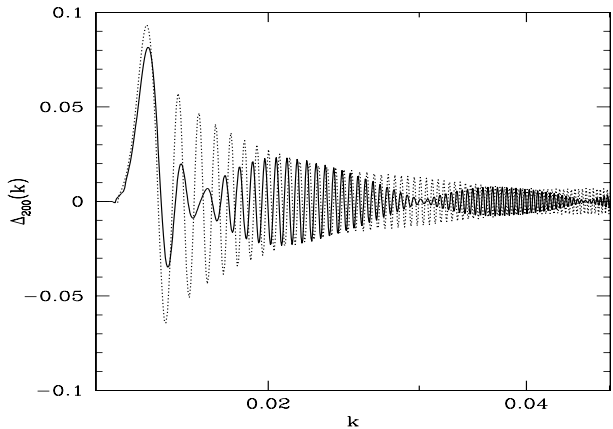,width=8.5cm,height=6cm}
\vskip +0.15in
\caption{The temperature transfer functions $\Delta_{200}(k)$ calculated
using the LOS method (dotted) and linear interpolation (solid). The
prominent feature is a beat of much lower frequency than the transfer
function. This is what limits the interpolation accuracy. (Note that
the actual code in DASh interpolates between the squares of the
transfer functions which decreases the effect of the beat).}
\label{beat}
\end{figure}

\subsection{Accuracy and Timing}

The LOS method is an extremely accurate method and is limited only by
the accuracy of the Bessel functions, which for flat models is
$0.1\%$, as was also demonstrated in \cite{seljak_code_comparison}.
Indeed this is the accuracy of all the transfer functions computed in
DASh. The approximation techniques introduce further errors, namely
interpolation error for all models and approximation error on large
scales for reionized models due to the approximate method used to include
reionization effects on large scales.

The interpolation error is due to an intrinsic $0.1\%$ error of the
transfer functions which is amplified by interpolation, and a beat
phenomenon which arises because the frequencies of the transfer
functions being interpolated are slightly offset from each other and
the target function. The intrinsic interpolation error could be
remedied if the transfer functions were calculated with an exact
Boltzmann code, though this would be at a cost in speed of
pre-computation. To illustrate the effect of the interpolation error
we plot the calculated LOS transfer function and the interpolated one
in Fig.~\ref{beat}. Note that the interpolation in Fig.~\ref{beat} is
between transfer functions and not their squares, as is actually done
in DASh, to dramatize the effect. Interpolating the squares of the
transfer functions reduces this error, at the expense of having to
interpolate $3\times10$ transfer functions (counting the temperature and
polarization for all auto- and cross-correlation modes) instead of
$2\times4$ (where the cross-correlation modes are calculated after interpolation).

\begin{figure*}
\center
\epsfig{file=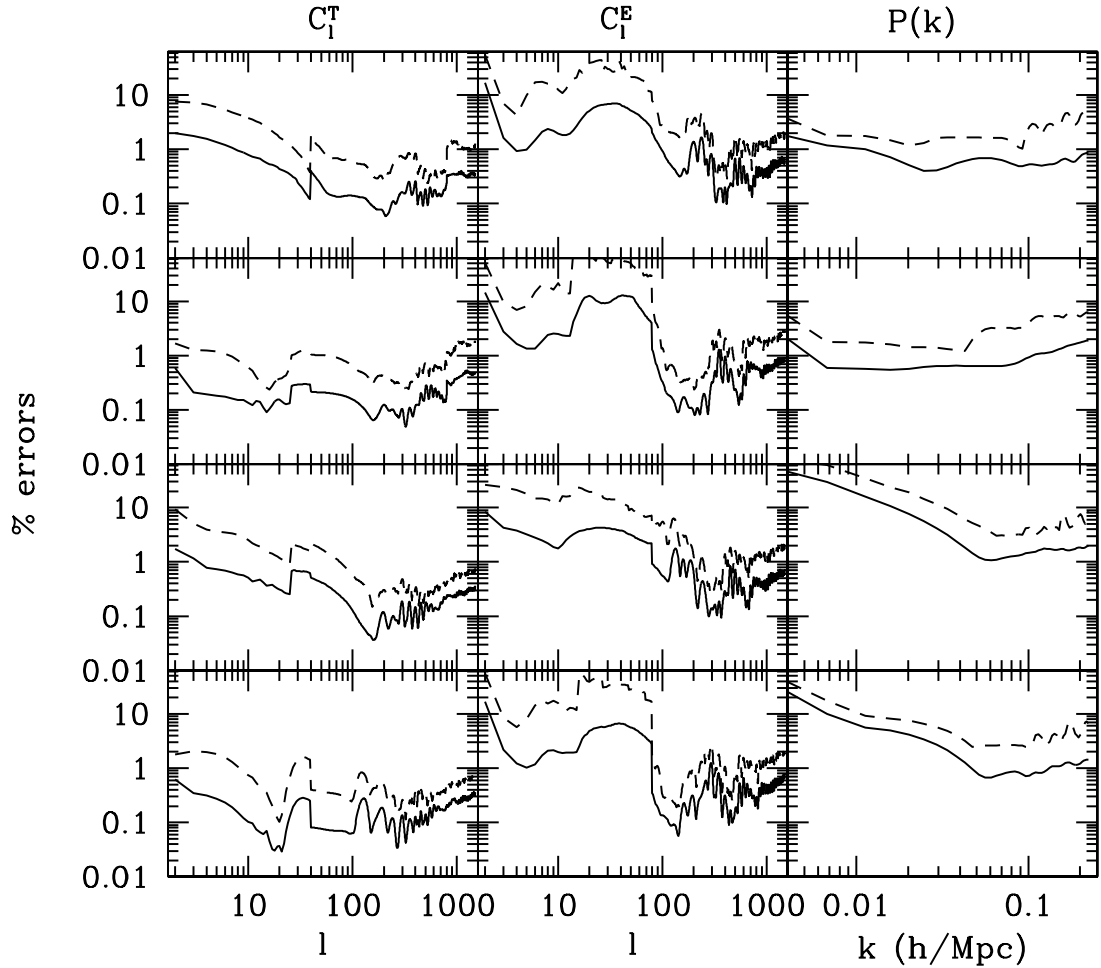,width=15cm,height=15cm}
\caption{
{\bf Maximum and RMS percentage differences between
the DASh interpolated spectra and the line-of-sight computed
spectra.}
Maximum (dashed) and RMS (solid) percentage differences of
DASh compared to the Line-Of-Sight method. From left to right we show
the CMB temperature spectrum, CMB polarization spectrum and matter
power spectrum $P(k)$. From top to bottom we show the adiabatic,
CDM isocurvature, neutrino isocurvature velocity and neutrino isocurvature
density modes. The errors for $C_\ell$ were calculated using 2048
models and for $P(k)$ using 512 models (we excluded variation of
$\tau$ for the matter power spectrum.}
\label{all_errors}
\end{figure*}

To check that the inclusion of additional initial conditions, and the
extension to compute the matter power spectrum, were sufficiently
accurate we tested the DASh code against $2048$ models sampled from a
grid in $\{\omega_b,\omega_c, \Omega_\Lambda,\tau\},$ in the case of
CMB temperature and polarization, and $512$ models sampled from a grid
in $\{\omega_b,\omega_c,\Omega_\Lambda\},$ in the case of the matter
power spectrum. We plot the maximum and root-mean-square (rms) errors
in Fig.~\ref{all_errors} for all auto-correlation mode spectra.  For
the CMB temperature spectrum (for which the most accurate measurements
exist) we see that the maximum error is less than 2\% over a wide
range in $\ell$, which is better than the WMAP measurement error. On
large scales ($\ell < 10$) the maximum error increases to 10\% for
some modes but this is still well below cosmic variance. 
 
We tested the effect of these numerical errors on the likelihood computed
for each model and found the measured difference to be negligibe.

The isocurvature-modified DASh described here took approximately
3--4 seconds on a 1GHz PentiumIII processor to compute the CMB temperature,
polarization, and cross-correlation spectra as well as the matter
power spectrum, for a single mode.
The full matrix of 4 modes and their cross-correlations required
approximately 15 seconds per model.

\end{document}